\documentclass[prd,aps,showpacs,%twocolumn,
11pt,%showpacs,
nofootinbib,%
tightenlines,%
notitlepage,%
eqsecnum%
]{revtex4-1}

\usepackage{amsmath,amssymb}
\usepackage{hyperref}
\usepackage{upgreek}
\usepackage[compat=1.1.0]{tikz-feynman}
\usepackage{xcolor}
\usepackage{soul}
\sethlcolor{yellow}

\newcommand{\be}{\begin{equation}}
\newcommand{\ee}{\end{equation}}
\newcommand{\bea}{\begin{eqnarray}}
\newcommand{\eea}{\end{eqnarray}}
\newcommand{\ba}{\begin{align}}
\newcommand{\ea}{\end{align}}

\newcommand{\RSS}{R_{\mathrm{D4}}}
\newcommand{\MKK}{M_{\mathrm{KK}}}

\newcommand{\gYM}{g_{\mathrm{YM}}}

\def\d{\text{d}}
\def\kk{\text{KK}}
\def\tr{\,\text{tr}\,}
\def\Tr{\,\text{Tr}\,}

\def\V{\mathcal{V}}

\def\xtau{\tau}

\def\Nc{N_c}

\def\PDG{\cite{Workman:2022ynf}}
\def\BPR{\cite{Brunner:2015oqa}}

\newif\ifmynotes
\mynotestrue
\definecolor{notetext}{rgb}{1,0,0}

\usepackage{color}
\begin{document}

\title{Spin-1 Glueballs in the Witten-Sakai-Sugimoto Model}
\author{Florian Hechenberger}
\author{Josef Leutgeb}
\author{Anton Rebhan}
\affiliation{Institut f\"ur Theoretische Physik, Technische Universit\"at Wien,
        Wiedner Hauptstrasse 8-10, A-1040 Vienna, Austria}

\date{\today}
\begin{abstract}
We consider the vector and the pseudovector glueball in the top-down holographic
model of large-$N_c$ QCD of Witten and their decays into ordinary mesons described
by the D8 brane construction due to Sakai and Sugimoto. At leading order,
the relevant interactions are determined exclusively by the Chern-Simons action
of the D8 branes and are thus rigidly connected to the chiral anomaly and the Wess-Zumino-Witten terms. As found in a previous study of the pseudovector glueball, which
we revisit and complete, the resulting decay widths are surprisingly large,
implying that both the pseudovector and the vector glueball are very broad resonances,
%whose dominant decay modes are $\pi\rho$ for the pseudovector and 
with a conspicuous dominance of decays into $a_1\rho$ and $K_1(1400) K^*$ in the case of the vector glueball. We also obtain a certain weak mixing of vector glueballs with ordinary vector mesons, but we conclude that it does not provide an explanation for the so-called $\rho\pi$ puzzle in charmonium decays.
% QCD model of Witten, Sakai and Sugimoto
% We consider the vector and pseudovector glueball in the top-down holographic construction by Sakai and Sugimoto and study dynamical quark effects through the backreaction of the D8-branes on the geometry. The action on the D8-brane introduces a small mixing between the vector glueball and singlet flavor gauge fields in the worldvolume theory. After computing the mixing angle for the three flavor case we explore its role in the $\rho\pi$ puzzle when extended to four flavors and find agreement with the bounds quoted in the literature. We also calculate various hadronic and radiative decays of the vector glueball and extend previous calculations of decays of the pseudovector glueball.
\end{abstract}
\maketitle  

\tableofcontents

\newpage

\section{Introduction and Summary}

In spite of extensive theoretical and experimental studies,
the status of glueballs in the hadron spectrum of QCD remains largely unsettled \cite{Klempt:2007cp,Crede:2008vw,Ochs:2013gi,Chen:2022asf}.
While the spectrum of glueballs as obtained in lattice QCD \cite{Bali:1993fb,Morningstar:1999rf,Chen:2005mg,Gregory:2012hu,Chen:2021dvn}
appears to be 
relatively stable when dynamical quarks are included, their interactions
and the amount of mixing with ordinary mesons are difficult to pin down, so that
no clear glueball state could be identified yet.

Lattice QCD indicates that the lightest glueball is a $J^{PC}=0^{++}$ scalar
with a mass between 1500 and 1800 MeV, but phenomenological studies
disagree \cite{Amsler:1995td,Close:2001ga,Close:2005vf,Lee:1999kv,Janowski:2014ppa,Cheng:2015iaa,Frere:2015xxa}
whether to identify it as a smaller or larger component
of the scalar-isoscalar mesons $f_0(1500)$, $f_0(1710)$, or a novel
$f_0(1770)$, or instead as a wide resonance distributed over several scalars
\cite{Klempt:2021nuf}.

The next lightest glueball is the $2^{++}$ tensor glueball associated
with the Pomeron \cite{Donnachie:2002en}, where lattice QCD indicates a mass around 2400 MeV, while
Pomeron physics favors a somewhat smaller mass,
followed by the $0^{-+}$ pseudoscalar around 2600 MeV, which
is expected to play a role in the chiral anomaly and the large $\eta'$ mass.

In this work we continue the studies of Ref.\ \cite{Hashimoto:2007ze,Brunner:2015oqa,Brunner:2015yha,Brunner:2015oga,Brunner:2018wbv,Leutgeb:2019lqu,Hechenberger:2023ljn} using the Witten-Sakai-Sugimoto (WSS) model \cite{Sakai:2004cn,Sakai:2005yt} to derive predictions for the
interactions of glueballs with ordinary mesons as well as their
radiative decays.
The Witten model \cite{Witten:1998zw} 
for low-energy large-$N_c$ QCD is based on a supersymmetry breaking
background geometry provided by an $N_c\gg1$ stack of circle compactified D4 branes in type-IIA supergravity, and it has a spectrum of spin-$0^{\pm+}$, spin-$1^{\pm-}$, and spin-$2^{++}$ glueballs with a mass hierarchy that is qualitatively in agreement with lattice findings
\cite{Constable:1999gb,Brower:2000rp}.
By adding stacks of $N_f\ll N_c$ D8 and anti-D8 probe branes,
Sakai and Sugimoto have succeeded in constructing a top-down holographic model
that provides a geometric model of nonabelian chiral symmetry breaking and reproduces numerous features of actual low-energy QCD qualitatively as well
as semi-quantitatively, typically with 10-30\% deviations, 
with a minimal number of free parameters. Because no further free parameters are involved to determine the interactions with
glueballs, the WSS model is also very predictive with respect to interactions between glueballs and ordinary mesons, which are treated as (approximately) unmixed in the 't Hooft limit
$g^2 N_c\gg1$, $N_f\ll N_c$, corresponding to a quenched approximation when we set $N_f=N_c=3$ in the end.

In Ref.~\cite{Hechenberger:2023ljn} we have recently revisited the predictions
of the WSS model for meson decays upon
including the $\eta'$ mass from the U(1)$_A$ anomaly and
adding a mass term for pseudoscalars induced by quark masses.
Besides extending the decay patterns of scalar and tensor glueballs
by radiative decay modes, we have also considered the pseudoscalar
glueball, which is represented by a Ramond-Ramond 1-form field and
whose interactions are determined by its anomaly-driven mixing with the $\eta_0$ meson.
The interactions of the latter are uniquely given by the Chern-Simons (CS)
term of the flavor branes, hence completely determined by the anomaly structure.
% entailing a width of $\sim 500$ MeV for the 
% pseudoscalar glueball when
% its mass is extrapolated to the QCD lattice prediction of $\sim 2600$ MeV.

In this paper, we extend the analysis to spin-1 glueballs, 
where the quenched lattice QCD simulation of Ref.~\cite{Chen:2005mg}
predicts masses around 3000 MeV for the pseudovector ($1^{+-}$) and around 3800 MeV
for the vector ($1^{--}$) glueball, which is reproduced well by the WSS model as far
as their ratio is concerned, while the overall scale is underestimated by about 30\%.
In the WSS model, the two spin-1 glueballs are represented by
the Kalb-Ramond tensor field in conjunction with a Ramond-Ramond 3-form field.
Their interactions with ordinary mesons are dominated by the unique
CS action of the D8 branes; they are thus tied to the structure of the
anomalous interactions of ordinary mesons. 
Moreover, through the Kalb-Ramond
field, the $1^{--}$ vector glueball mixes with the singlet component of 
ordinary vector mesons, which is interesting with regard to
the proposal \cite{Hou:1982kh,Brodsky:1987bb,Chan:1999px} that mixing with vector glueballs could
explain the so-called $\rho\pi$ puzzle in charmonium decays \cite{Mo:2006cy},
which consists of a surprisingly strong suppression of $\rho\pi$ and $K^* K$
in the decay of $\psi(2S)$ compared to $\psi(1S)=J/\psi$.
However, in the WSS model the decay pattern of the vector glueball
turns out to
have a strong enhancement in the $a_1\rho$, $K_1 K^*$, and $f_1\omega$ channels,
which are not seen in any of the $\psi(nS)$ decays. 
The results of the WSS model thus do not support an explanation of the
charmonium $\rho\pi$ puzzle
through vector glueball admixtures.
% These results therefore
% do not provide a possible explanation, even if the $\psi$ states 
% in question
% (which themselves can hardly
% be described by the WSS model) would have a very different mass-dependent mixing with
% vector glueballs.

The couplings and decay patterns of vector and pseudovector glueballs are also of interest
with regard to the physics of the Odderon \cite{Donnachie:2002en,Ewerz:2003xi}, which recently has been claimed to have been discovered in joint experiments by the TOTEM and D0 collaborations \cite{D0:2020tig}. Brower et al.\ \cite{Brower:2008cy}
have argued that in holographic QCD Odderons appear naturally as the Reggeized 
Kalb-Ramond modes in the
Neveu-Schwarz sector of closed string theory, which contains both
vector and pseudovector glueball modes whose interactions with ordinary
hadrons are fixed in the WSS model without any additional free parameters.

However, as found in the previous study of the decays of the pseudovector
glueball in Ref.~\cite{Brunner:2018wbv}, which we revisit and complete,
the decay widths obtained in the WSS model are very large, making
both spin-1 glueballs difficult to discover, albeit the peculiar
decay pattern of the vector glueball may be helpful in this respect.

This paper is organized as follows.
In Sec.\ \ref{sec:review} we recapitulate the WSS model
as used in \cite{Hechenberger:2023ljn}, but expanded to include all form fields relevant for spin-1 glueballs.
In Sec.\ \ref{sec:vgb} we derive the bulk mode function of the vector glueball and describe its effects on the hadronic modes on the flavor branes, followed by a systematic evaluation of the hadronic and radiative decay modes, closing with a discussion of the implications for the $\rho\pi$ puzzle in $J/\psi$ and $\psi'$ decays. In Sec.~\ref{sec:pvgb}, we consider the pseudovector glueball, revisiting and completing the previous work of Ref.~\cite{Brunner:2018wbv}. Sec.~\ref{sec:concl} contains our conclusions and comments on phenomenological consequences.

\section{Quick review of the Witten-Sakai-Sugimoto model}
\label{sec:review}

The 10-dimensional background geometry corresponding to an
$N_c\gg1$ stack of D4 branes compactified with supersymmetry breaking boundary conditions in the circular fourth spatial coordinate $x^4\equiv\tau$,
\begin{align}
\xtau\simeq \xtau+\delta\xtau=\xtau+2\pi M_{\kk}^{-1},
%\quad & M_{\kk}=\frac{3}{2}\frac{U_\kk^{1/2}}{\RSS^{3/2}},
\end{align}
is given by the metric
\begin{align}
  & \d s^2 = \left( \frac{U}{\RSS} \right)^{3/2} \left[\eta_{\mu\nu} \d x^\mu \d x^\nu + f(U) \d \xtau^2\right]+\left(\frac{\RSS}{U}\right)^{3/2} \left[\frac{\d U^{2}}{f(U)} + U^2 \d\Omega_4^2\right],\nonumber \\
  & e^{\phi} = g_{s}\left(\frac{U}{\RSS}\right)^{3/4},\qquad F_{4}=\d C_{3}=\frac{(2\pi l_s)^3\Nc}{V_{4}}\epsilon_{4},\qquad f(U)=1-\frac{U_\kk^3}{U^{3}},\label{eq:background}
\end{align}    
with dilaton $\phi$ and Ramond-Ramond three-form field\footnote{Using standard string-theory conventions \cite{Polchinski1998} for
the normalization of Ramond-Ramond fields
rather than the rescaled version of Ref.~\cite{Sakai:2004cn}.}
$C_{3}$.
Here $x^{\mu}$, $\mu=0,1,2,3$, are the coordinates in the flat four-dimensional directions, $U$ is the radial holographic direction, where
regularity at $U=U_\kk$ fixes
\begin{equation}
    M_{\kk}=\frac{3}{2}\frac{U_\kk^{1/2}}{\RSS^{3/2}};
\end{equation}
the radius
$\RSS$ is related to the string coupling $g_{s}$ and the string length
$l_{s}$ through $\RSS^{3}=\pi g_{s}N_{c}l_{s}^{3}$,
and the 't Hooft coupling of the dual four-dimensional Yang-Mills
theory that arises after Kaluza-Klein reduction is given by
\be\label{gYMNc}
\lambda=\gYM^2 N_c=\frac{g_5^2}{\delta\xtau}N_c=2\pi g_s l_s\MKK N_c.
\ee

% It is obtained by placing a stack of $N_f\times \overline{N}_f$ probe branes in the doubly Wick rotated black D4-brane background proposed in \cite{Witten:1998zw} as a dual to $\mathcal{N}=4$ super Yang-Mills theory at large $N_c$. Conformal symmetry and supersymmetry are both broken by appropriate boundary conditions for the fermion sector and hence masses for the adjoint scalars are generated at the one-loop level. Explicitly, the background in which the probe branes are placed is given by
% \begin{align}
%   & \d s^2 = \left( \frac{U}{\RSS} \right)^{3/2} \left[\eta_{\mu\nu} \d x^\mu \d x^\nu + f(U) \d \xtau^2\right]+\left(\frac{\RSS}{U}\right)^{3/2} \left[\frac{\d U^{2}}{f(U)} + U^2 \d\Omega_4^2\right],\nonumber \\
%   & e^{\phi} = g_{s}\left(\frac{U}{\RSS}\right)^{3/4},\qquad F_{4}=\d C_{3}=\frac{2\pi\Nc}{V_{4}}\epsilon_{4},\qquad f(U)=1-\frac{U_\kk^3}{U^{3}},\label{eq:background}
% \end{align}    
% \hl{Faktor von F4 falsch?}
% with a non-constant dilaton $\phi$ and Ramond-Ramond three-form field $C_{3}$ obtained by the flux of the $N_c$ D4 branes. It is a solution
This is a solution in type IIA supergravity, whose bosonic part
reads \cite{Polchinski1998}
% \begin{widetext}
% \begin{equation}
% S_{\text{grav}}= \frac{1}{2\kappa_{10}^2}\int\d^{10}x\sqrt{-g}\left[e^{-2\phi}\left(R+4\left(\nabla\phi\right)^{2}\right)-\frac{(2\pi)^4l_s^6}{2}\left|F_{4}\right|^{2}\right].\label{eq:bulkAction}
% \end{equation}    
% \end{widetext}
\begin{equation}
    \begin{split}
    S_{IIA}&=S_{NS}+S_{R}+S_{CS},\\
        S_{NS}&=\frac{1}{2\kappa_{10}^2}\int\d^{10}x\sqrt{-g}e^{-2\phi}\left(R+4\nabla_M\phi\nabla^M\phi-\frac{1}{2}|H_3|^2\right),\\
        S_R&=\frac{1}{2\kappa_{10}^2}\int\d^{10}x\sqrt{-g}\left(-\frac{1}{2}|F_2|^2-\frac{1}{2}|\tilde{F}_4|^2\right),\\
        S_{CS}&=-\frac{1}{2\kappa_{10}^2}\int\d^{10}x\frac{1}{2}B_2\wedge F_4\wedge F_4,
    \end{split}
    \label{eq:IIaAction}
\end{equation}
where
\begin{equation}
    \begin{split}
        F_2=dC_1,&\quad         F_4=dC_3,\\
        \Tilde{F}_4=F_4-&C_1\wedge H_3,\quad H_3=dB_2.
    \end{split}
    \label{eq:IIAmFieldRed}
\end{equation}
%
% The world volume coordinates of the $N_c$ D4-branes are given by
% $x^{\mu}$, $\mu=0,1,2,3$ and a compactified $\xtau$-coordinate on which the supersymmetry breaking boundary conditions are imposed with period
% \begin{align}
% \xtau\simeq \xtau+\delta\xtau=\xtau+2\pi M_{\kk}^{-1},\quad & M_{\kk}=\frac{3}{2}\left(\frac{U_\kk}{\RSS^3}\right)^{1/2},
% \end{align}
% to avoid a conical singularity at $U=U_{\kk}$. The transverse coordinates are given by the radial holographic direction $U$ and the unit $S_4$ with line element $\d\Omega_4^2$, volume form $\epsilon_4$  and volume $V_{4}=8\pi^{2}/3$. 
% The relation between $\RSS$, the string coupling $g_{s}$ and the string length
% $l_{s}$ is fixed by the $F_4$ flux to $\RSS^{3}=\pi g_{s}N_{c}l_{s}^{3},$ this fixes the 't Hooft coupling on the field theory side to
% \be\label{gYMNc}
% \lambda=\gYM^2 N_c=\frac{g_5^2}{\delta\xtau}N_c=2\pi g_s l_s\MKK N_c.
% \ee
%
The probe ($N_f\ll N_c$) D8 and $\overline{\text{D8}}$-branes extend along $x^{\mu}$,
$U$, $S^{4}$ and are located in an antipodal configuration on the $\xtau$-circle, joining smoothly at $U_\kk$, thereby
realizing spontaneous $\mathrm{U}_L(N_f)\times \mathrm{U}_R(N_f)$ breaking. 

% They are placed antipodally on the $\xtau$-circle
% to join at $U_\kk$. In adopting the probe approximation, i.e. $\Nc\gg N_{f}$
% for the $N_{f}$ D8-branes, one can ignore backreactions from the D8-branes
% to the D4-brane background. The gauge fields on the D8-branes, which are dual to left and right chiral quark currents separated in the Kaluza-Klein ($\tau$) direction, are
% governed at leading order by a Dirac-Born-Infeld (DBI) plus Chern-Simons
% (CS)  action
% \begin{align}
% S_{\text{DBI}}= & -T_{8}\int\d^{9}x e^{-\phi}\Tr \sqrt{-\det\left(g_{MN}+2\pi\alpha^{\prime}F_{MN}\right)},\nonumber \\
% S_{\text{CS}}= & T_{8}\int_{D8}C\wedge\Tr\left[\exp\left\{ \frac{F}{2\pi}\right\} \right]\sqrt{\hat{A}(\mathcal{R})},\label{eq:9dAction}
% \end{align}
The action for the flavor D8-branes is given by the sum of the DBI action and the Chern-Simons action
\begin{equation}
    \begin{split}
        S_{DBI}^{\text{D8}}&=-T_8\int\d^9xe^{-\phi}\Tr \sqrt{-\det\left(g_{MN}+2\pi\alpha' {F}_{MN}+B_{MN}\right)},\\
        S_{CS}^{\text{D8}}&=T_8\sum_p\int_{\text{D8}}\sqrt{\hat{\mathcal{A}}(\mathcal{R})}\Tr \exp\left(2\pi\alpha' {F}+B\right)\wedge C_p,
    \end{split}
    \label{eq:D8Action}
\end{equation}
with ${F}$ the nonabelian flavor field strength and $\hat{A}(\mathcal{R})$ being the A-roof genus \cite{Green:1996dd,Polchinski1998}. 
The sum in the Chern-Simons term is a formal sum over the p-form gauge fields in the Ramond-Ramond sector of the theory.  
% Note that Eq.\eqref{eq:IIaAction} is stricly speaking not valid when the calculations are taken beyond the probe limit of the D8-branes. This is due to the appearance of the $C_9$ potential which is sourced by the D8-branes, but absent in the conventional type IIA supergravity action in Eq.\eqref{eq:IIaAction}. One can show that this potential leads to an additional mass term in the action for the Kalb-Ramond field $B_2$ and introduces a mixing between $F_2$ and $B_2$. \footnote{See \cite{Burrington:2007qd} for a discussion in the $D4/D8$-brane system.} A generalized action with some redefined field strengths resulting in a consistent set of equations of motion has been obtained by Romans in \cite{Romans:1985tz}.  However, one can check that these modifications do not contribute to linear order in $N_f/N_c$ for the fluctuations of the  closed string sector under consideration here. In particular the $1^{\pm -}$ fluctuations do not receive mass corrections via a Higgs type mechanism with $0^{-+}$, when the spectra are truncated at the lowest Kaluza-Klein (KK) mode in $\xtau$. In the following we will thus work with Eq.\eqref{eq:IIaAction} and restrict our calculations to the lowest KK modes. This is also the case when working with smeared branes as in \cite{Bigazzi:2014qsa}, which we will use later on for our estimates of dynamical quark effects.\\

Following \cite{Sakai:2004cn,Sakai:2005yt}, the spectrum on the joined D8 and $\overline{\text{D8}}$-brane is truncated to include only $SO(5)$ invariant states. To this end, and to quadratic order, the DBI action in Eq.\eqref{eq:5dDBI} reduces to\footnote{Note that in \eqref{eq:5dDBI} one uses the Minkowski metric $\eta_{\mu\nu}$, in the mostly plus
convention, to contract the four-dimensional spacetime indices.}
\begin{equation}
S^\text{D8}_\text{DBI} = -\kappa \int  \d^4x\, \d z\, {\Tr}\left[ \frac{1}{2} K^{-1/3} F_{\mu\nu}^2 + M_\kk ^2 K F_{\mu z}^2 \right],\label{eq:5dDBI}
\end{equation}
with
\be
\kappa\equiv\frac{\lambda N_{c}}{216\pi^{3}},\qquad K(z)\equiv1+z^{2}={U^{3}}/{U_{\kk}^{3}},
\ee
where $z$ runs from $-\infty$ to $+\infty$ along the
joined D8 branes.

Performing a Kalzua-Klein(KK) decomposition for the five-dimensional flavor gauge fields
\begin{align}
        &A_{\mu}(x^\mu,z)=\sum_{n=1}^{\infty}B_{\mu}^{\left(n\right)}(x^{\mu})\psi_{n}(z)\nonumber\\
&A_{z}(x^\mu,z)= \sum_{n=0}^{\infty}\varphi^{\left(n\right)}(x^{\mu})\phi_{n}(z),
    \label{eq:separationAnsatz} 
\end{align}
yields a tower of massive vector and axial vector mesons
corresponding to odd and even mode numbers $n$ with
even and odd $z$-parity, respectively (see our previous
paper \cite{Hechenberger:2023ljn} for further details):
\bea
&v_{\mu}^{n}\equiv B_{\mu}^{\left(2n-1\right)},\qquad a_{\mu}^{n}\equiv B_{\mu}^{\left(2n\right)},
\eea
Identifying the lightest vector mode with the $\rho$ meson
fixes $M_\kk=949$  MeV \cite{Sakai:2004cn,Sakai:2005yt}, corresponding to $m_\rho=776.4$ MeV.

The scalar fields $\varphi^{(n)}$ can be
absorbed by the fields $B_\mu^{(n)}$ except for $\varphi^{(0)}$
which corresponds to the massless pseudoscalar Goldstone multiplet of the broken chiral symmetry,
\be\label{eq:Ux}
U(x)=e^{i\Pi^a(x)\lambda^a/f_\pi}=\mathrm P\,\exp i\int_{-\infty}^\infty \d z\, A_z(z,x),
\ee
with the Gell-Mann matrices $\lambda^a=2T^a$ and including the singlet term $\lambda^0=\sqrt{2/N_f}\mathbf{1}$.

To fix the 't Hooft coupling $\lambda$ we use the resulting pion decay constant 
\be\label{fpi2}
f_\pi^2=\frac{\lambda N_c\MKK^2}{54\pi^4}
\ee
to get $\lambda\approx 16.63$ from $f_\pi\approx 92.4$ MeV. To obtain an error estimate and following \BPR ~we  shall also consider the smaller value $\lambda\approx 12.55$ obtained by matching the large-$N_c$ lattice result for the string tension obtained in Ref.~\cite{Bali:2013kia}.

The non-normalizable modes of the flavor gauge field $A_\mu$ can be used to introduce the photon field as an external source via \cite{Sakai:2005yt} 
\be
\lim_{z\to\pm\infty}A_\mu(x,z)=A_{L,R\mu}(x)= eQA_{\mu}^{\text{em}}(x),
\ee
with the quark charge matrix $Q$ for $N_f=3$ given by
\begin{eqnarray}
Q & =\frac{1}{3}\left(\begin{array}{ccc}
2\\
 & -1\\
 &  & -1
\end{array}\right),
\end{eqnarray}
where $e$ is the electromagnetic charge. As reviewed
in our previous paper \cite{Hechenberger:2023ljn}, vector meson dominance (VMD) arises because the photon field couples exclusively through mixing with the tower of vector mesons.
For on-shell photons, the corresponding holographic wave function
entering the overlap integrals with the mode functions of
hadronic fields
reduces to unity; off-shell photons involve nontrivial
bulk-to-boundary propagators.

For $N_f=3$, which we shall consider in the following,
we also take into account that in the WSS model the U(1)$_A$ flavor symmetry is broken by an anomalous contribution of
order $1/N_c$ due to the $C_1$ Ramond-Ramond field, which gives rise to a 
Witten-Veneziano \cite{Witten:1979vv,Veneziano:1979ec} mass term for the singlet $\eta_0$ pseudoscalar
with \cite{Sakai:2004cn,Leutgeb:2019lqu}
\be\label{mWV2}
m_{0}^2=\frac{N_f}{27\pi^2 N_c}\lambda^2\MKK^2.
\ee
For $N_f=N_c=3$, one has
$m_{0}=967\dots730$ MeV for $\lambda=16.63\dots12.55$, which
is indeed a phenomenologically interesting ballpark when
finite quark masses are added to the model by the
addition of an effective Lagrangian
\be\label{Lm}
\begin{split}
    &\mathcal{L}_m^{\mathcal M} \propto\Tr\left(\mathcal M\,U(x)+h.c.\right),\\
    &\mathcal M={\rm diag}(m_u,m_d,m_s),  
\end{split}
\ee
which can be motivated by either worldsheet instantons \cite{Aharony:2008an,Hashimoto:2008sr}
or nonnormalizable modes of additional bifundamental fields corresponding to open-string tachyons \cite{0708.2839,Dhar:2008um,McNees:2008km,Niarchos:2010ki}.

Assuming for simplicity isospin symmetry, $m_u=m_d=\hat m$,
this leads to masses \cite{Brunner:2015oga,Hechenberger:2023ljn}
%     \be
% m^2_{\eta,\eta'}=\frac12 m_0^2+m_K^2\mp\sqrt{\frac{m_0^4}{4}-\frac13 m_0^2(m_K^2-m_\pi^2)+(m_K^2-m_\pi^2)^2}
% \ee
% for the mass eigenstates
% \bea
% \eta &=& \eta_8 \cos\theta_P - \eta_0 \sin\theta_P\nn\\
% \eta'&=& \eta_8 \sin\theta_P + \eta_0 \cos\theta_P,
% \eea
% with mixing angle
% \be\label{thetaP}
% \theta_P=\frac12\arctan\frac{2\sqrt2}{1-\frac32 {m_0^2}/({m_K^2-m_\pi^2})}.
% \ee
% Using
% $m_\pi^2=m_{\pi_0}^2\approx (135 {\rm MeV})^2$ and
% %$m_K^2=(495{\rm MeV})^2$,
% \be
% m_K^2=\frac12(m_{K_\pm}^2+m_{K_0}^2)-
% \frac12(m_{\pi_\pm}^2-m_{\pi_0}^2)\approx (495{\rm MeV})^2
% \ee
% as isospin symmetric parameters,
% the WSS result $m_0\approx 967\dots730$ MeV for $\lambda=16.63\dots12.55$
% leads to $\theta_P\approx -14^\circ\dots-24^\circ$
% and 
$m_\eta\approx 520\dots 470$, $m_{\eta'}\approx 1080\dots 890$ MeV and mixing angles $\theta_P\approx -14^\circ\dots-24^\circ$ for $\lambda=16.63\dots12.55$.

In the following we shall consider this range of mixing angles in conjunction with the variation of $\lambda$, but we shall fix $m_\eta$ and $m_{\eta'}$
to their experimental values when evaluating phase space integrals.

Vector mesons remain unchanged by this introduction
of quark masses. In the following we shall keep the (chiral) results for their couplings, but we will
raise the masses of $\omega$ and $\phi$ mesons
to their experimental values in phase space integrals,
assuming ideal mixing.

In the WSS model, the axial vector meson $a_1$ is
predicted with mass 1186.5 MeV, very close to the experimental result of 1230(40) MeV. For the
remaining axial vector mesons we again keep the
chiral results for their couplings, but introduce
phenomenological masses and mixing angles in phase space
integrals. Here we use a mixing angle of $\theta_f=20.4^\circ$ for $f_1$ and $f_1'$ mesons
in
\begin{align}\label{f1mixing}
    |f_1(1285) \rangle&=\cos \theta_f|\bar n n\rangle-\sin \theta_f|\bar s s\rangle ,\nonumber\\
    |f_1(1420) \rangle&=\sin \theta_f |\bar n n\rangle +\cos \theta_f |\bar s s\rangle .
\end{align}
The physical strange axial vector mesons $K_1(1270)$ and $K_1(1400)$ are mixtures of $K_{1A}$ $(1^{++})$ and the excited axial vector meson $K_{1B}$ ($1^{+-}) $\cite{Suzuki:1993yc}.
Because in the WSS model,
there is no $1^{+-}$ nonet of ordinary mesons,
only $K_{1A}$ is present, which couples to the
physical $K_1$ mesons according to their mixing defined by
\begin{equation}
    |K_{1A}\rangle =  \cos\theta_K |K_1(1400)\rangle
    + \sin\theta_K |K_1(1270)\rangle.
\end{equation}
In \cite{Suzuki:1993yc,Divotgey:2013jba} the favored mixing angle is quoted as $|\theta_K|\approx 33^\circ$, which we adopt in the following.

Encouragingly, the WSS model predicts rather well the ballpark
of several hadronic decays such as $\rho\to\pi\pi$, $\omega\to\pi\pi\pi$, $a_1\to\rho\pi$, and also various radiative decays,
see Ref.~\cite{Sakai:2004cn,Sakai:2005yt,Brunner:2015oqa,Hechenberger:2023ljn}.

\section{The Vector Glueball in the WSS}
\label{sec:vgb}

% \0{Keep exponential form? Not really necessary if we don't include the backreaction}
The mass spectra for the spin-1 fluctuations in the M-theory lift of the Witten model were first obtained in \cite{Brower:2000rp} by considering the fluctuations of $A_{MNO}$ and $A_{MN11}$. In the 10D string frame, these fluctuations translate to $C_3$ and $B_2$, respectively. %Since we want to consider the effect of the backreaction of the D8-branes on the geometry as well and no known lift of this system to 11D IIA supergravity is known, we will work in the 10D string frame instead. 

Treating contributions stemming from the D8-branes as perturbations later on, the relevant %linearized 
field equations are obtained by varying Eq.\eqref{eq:IIaAction} with respect to $B_2$ and $C_3$
% \begin{equation}
%     \begin{split}
%         \nabla_O\left(e^{-2\phi}H^{OMN}+C_P \Tilde{F}^{OPMN}\right)-\frac{1}{2!\cdot(4!)^2\sqrt{-g}}\epsilon^{MNO_1\hdots O_8}F_{O_1\hdots O_4}F_{O_5\hdots O_8}&=0,\\
%         \nabla_P\Tilde{F}^{PMNO}-\frac{1}{3!\cdot 4!\sqrt{-g}}\epsilon^{MNOP_1\hdots P_7}H_{P_1P_2P_3}F_{P_4\hdots P_7}&=0,\\
%     \end{split}
% \end{equation}
\begin{equation}
    \begin{split}
        \nabla_O\left(e^{-2\phi}H^{OMN}\right)-\frac{1}{2!\cdot(4!)^2\sqrt{-g}}\epsilon^{MNO_1\hdots O_8}F_{O_1\hdots O_4}F_{O_5\hdots O_8}&=0,\\
        \nabla_P F^{PMNO}-\frac{1}{3!\cdot 4!\sqrt{-g}}\epsilon^{MNOP_1\hdots P_7}H_{P_1P_2P_3}F_{P_4\hdots P_7}&=0.
    \end{split}
    \label{eq:IIaEoms}
\end{equation}
% where we took into account that only $C_3$ is sourced by the D4 branes and hence $\tilde{F}_4=F_4$. The two additional contributions originating from the DBI action result in a mixing term of order $\mathcal{O}(\sqrt{N_f/N_c})$, and a $\mathcal{O}(N_f/N_c)$ mass correction. For completeness, and to fix our conventions, we also give the components of $F_4$\footnote{Recall that we use the usual definition for the form fields found in the string theory literature. In \cite{Sakai:2004cn} the form fields are rescaled according to $C_p=(2\pi)^{p-1}l_s^p C_p^{SS}$.}
% \begin{equation}
%     F_{\alpha\beta\gamma\delta}=\frac{3 R^3}{g_s}\sqrt{\hat{g}_{S_4}}\epsilon_{\alpha\beta\gamma\delta}
% \end{equation}

\subsection{Ansatz, normalization and equations of motion}
In \cite{Brower:2000rp} the $1^{--}$ vector glueball mode is obtained from the $A_{\mu\nu\tau}$ and $A_{\mu r 11}$ components of the 11D gauge field $A_3$ which translates to $C_{\mu\nu\tau}$ and $B_{\mu u}$ in the 10D string frame. Note that including the $B_{\mu u}$ fluctuation is necessary to obtain a consistent solution of the equation of motion since these two fluctuations are tied by a topological mass term. Starting from the ansatz
\begin{equation}
    C_{\mu\nu\tau}=\frac{a(u)}{g_s}\tilde{C}_{\mu\nu}(x^\mu),\quad B_{\mu u}=\frac{3}{2\Box}\frac{u^2}{u^3-1}a(u)\eta_{\mu\kappa}\epsilon^{\kappa\nu\rho\sigma}\partial_\nu \tilde{C}_{\rho\sigma}(x^\mu), 
    \label{eq:vectorAnsatz}
\end{equation}
% \begin{equation}
%     C_{\mu\nu\tau}=\frac{a(r)}{g_s}\tilde{C}_{\mu\nu}(x^\mu),\quad B_{\mu r}=\frac{3}{2\Box}e^{4\lambda}a(r)\eta_{\mu\kappa}\epsilon^{\kappa\nu\rho\sigma}\partial_\nu \tilde{C}_{\rho\sigma}(x^\mu), 
%     %\label{eq:vectorAnsatz}
% \end{equation}
and neglecting backreactions from the DBI action, we obtain the mode equation for the vector glueball
% \begin{equation}
%     a''(r)+\left(\varphi'-\tilde{\lambda}'+4\nu'\right)a'(r)+\left(M^2\frac{R^3}{U_\kk}e^{-2\lambda-2\varphi}-9e^{2\phi-2\varphi-8\nu}\right)a(r)=0.
%     \label{eq:vectorModeEOM}
% \end{equation}
%\hl{F:new:}
\begin{equation}\label{eq:vectorModeEOM}
    a''(u)+a'(u)/u+a(u)\left(M^2\frac{R^3}{U_\kk(u^3-1)}-\frac{9u}{u^3-1}\right)=0,\quad u=U/U_\kk.
\end{equation}
% \hl{Transform to standard Witten model coordinates and rescale to match eom in} \cite{Brower:2000rp}
The relation to the notation used in \cite{Brower:2000rp} is
\begin{equation}
    a(u)=\sqrt{r^6/r_\kk^6-1}M_4(r),\quad u^3=r^6/r_\kk^6,
\end{equation}
and when using coordinates $z$ along the D8 branes we have
$a(z)=z M_4(z)$.

% By rescaling the radial wave function 
% \begin{equation}
%     a(r)\to \frac{1}{\sqrt{e^{3r}-1}}M_4(r)
% \end{equation}
% we match the mode equation previously in \cite{Brower:2000rp}.
By imposing the boundary conditions $M_4'(U_\kk)=1$ and $M_4(\infty)=0$ we obtain the mass spectrum $M_V^2=\lambda_V M_\kk^2$ with the first three eigenvalues given by $\lambda_V=\{9.22721, 15.9535,24.1552\}$. The lowest eigenvalue corresponds to the mass of $M_V=2883$ MeV 
which is % slightly heavier then the $1^{+-}$ glueball but still 
below the (quenched) lattice result of $\approx 3850$ MeV\cite{Morningstar:1999rf,Chen:2005mg}. 

To fix the normalization we induce the fluctuations \eqref{eq:vectorAnsatz} in \eqref{eq:IIaAction} and utilize the equation of motion \eqref{eq:vectorModeEOM} to get
% \begin{equation}
%     \mathcal{L}_V^{(2)}dr=-\frac{1}{2\kappa_{10}^2}\frac{1}{4g_s^2}\frac{R^6}{U_\kk}\sqrt{\hat{g}_{S_4}}e^{-2\lambda-\Tilde{\lambda}-\varphi+4\nu}a(r)^2 \tilde{C}_{\mu\nu}\left(M^2-\Box\right)\tilde{C}_{\mu\nu}dr,
% \end{equation}
% \hl{F: new, transformed to z to match coordinates of brane action}
% \begin{equation}
%     \mathcal{L}_V^{(2)}dz=-\frac{1}{2\kappa_{10}^2}\frac{1}{4g_s^2}\frac{R^6}{U_\kk}\sqrt{\hat{g}_{S_4}}\frac{2}{3}\frac{1}{(1+z^2)^{1/3}}a(z)^2 \tilde{C}_{\mu\nu}\left(M^2-\Box\right)\tilde{C}_{\mu\nu}dz,
% \end{equation}
% \hl{F: new, transformed to u}
\begin{equation}
    \mathcal{L}_V^{(2)}=-\frac{1}{2\kappa_{10}^2}\frac{1}{4g_s^2}\frac{R^6}{U_\kk}\sqrt{\hat{g}_{S_4}}\frac{u}{u^3-1}a(u)^2 \tilde{C}_{\mu\nu}\left(M^2-\Box\right)\tilde{C}_{\mu\nu}.
\end{equation}
%from which we can infer the normalization condition
Requiring a kinetic term with canonical normalization
after integrating over the holographic coordinate, the $S_4$,
and the $S_1$, we set
$a(u)\to a(u)/\mathcal{N}_V$ with
% \begin{equation}
%     \int\d r\frac{1}{2\kappa_{10}^2}\frac{2\pi}{M_\kk}V_4\frac{R^6}{U_\kk g_s^2}e^{-2\lambda-\Tilde{\lambda}-\varphi+4\nu}\mathcal{N}_V^{-2}a(r)^2=1,
% \end{equation}
% \hl{F: new in z and u}
% \begin{equation}
%     \int\d z\frac{1}{2\kappa_{10}^2}\frac{2\pi}{M_\kk}V_4\frac{R^6}{U_\kk g_s^2}\frac{2}{3}\frac{1}{(1+z^2)^{1/3}}\mathcal{N}_V^{-2}a(z)^2=1,
% \end{equation}
\begin{equation}
    \int\d u\frac{1}{2\kappa_{10}^2}\frac{2\pi}{M_\kk}V_4\frac{R^6}{U_\kk g_s^2}\frac{u}{u^3-1}\mathcal{N}_V^{-2}a(u)^2=1,
\end{equation}
% where we integrated over the $S_4$ with volume $V_4=\frac{8\pi^2}{3}$ and radius $R=L/2$ and the $S_1$ with radius $M_\kk^{-1}$. Transforming to the usual WSS coordinate this translates to
%
% \begin{equation}
%     \int\d U \frac{U U_\kk}{U_\kk^3-U^3}\frac{1}{2\kappa_{10}^2}\frac{2\pi}{M_\kk}M_4\frac{R^6}{U_\kk g_s^2}\mathcal{N}_V^{-2}a(U)^2=1.
% \end{equation}
%
%From this we obtain the normalization
% \begin{equation}
%     \mathcal{N}_V^2=\frac{3}{16}\frac{\lambda N_c^2}{(2\pi)^2}\frac{1}{M_\kk^4 R^6}\int\d u\frac{u a(u)^2}{1-u^3},\quad u=\frac{U}{U_\kk}.
% \end{equation}
%
% If we work with the rescaling
% \begin{equation}
%     a(r)\to \frac{1}{\sqrt{e^{3r}-1}}M_4(r)
% \end{equation}
% this changes to
leading to
\begin{equation}
    \mathcal{N}_V^2=\frac{3}{16}\frac{\lambda N_c^2}{(2\pi)^2M_\kk^4 R^6}\int_1^\infty\d u u M_4(u)^2,\quad u=\frac{U}{U_\kk},
\end{equation}
% where we rescaled $a(r)=\frac{1}{\sqrt{e^{3r}-1}}M_4(r)$ to obtain
with
\begin{equation}
    \mathcal{N}_V=0.0142218 \frac{\sqrt{\lambda} N_c}{M_\kk^2 R^3}
\end{equation}
for the ground-state vector glueball.

%\subsubsection{Z parity of the radial modes}
When considering interactions with modes on the flavor branes, 
the integration variable $z$ covers the holographic radial coordinate twice.
The glueball modes are all even under $z$-parity. However the rescaling employed above corresponds to $a(z)=z M_4(z)$ and thus $M_4(z)$ has odd parity on the joint flavor branes.

% For reference we list the z parity of all the fields involved in our calculations
% \begin{equation}
%     \begin{split}
%         \phi_0(z)=\phi_0(-z), \ \psi_{2m-1}(z)=\psi_{2m-1}&(-z), \ \psi_m(z)=-\psi_m(-z),\\
%         M_4(z)=-M_4(-z)&,\ N_4(z)=N_4(-z) ,
%     \end{split}
% \end{equation}
% where $N_4(z)$ is the radial mode of the pseudovector glueball.

\subsection{Bilinear corrections due to the DBI action}

Because the Kalb-Ramond field couples directly to the flavor branes
through the DBI action, the latter gives rise to bilinear terms
involving the vector glueball field and the singlet component of the
vector meson field.

\subsubsection{Mass correction}

Integrating over the holographic direction and the $S^4$,
the DBI action gives rise to an additional mass term for the vector glueball
proportional to $N_f/N_c$, given by
\begin{equation}
    \begin{split}
        S_{DBI}&=-T_8\tr\int\d^9x e^{-\phi}\sqrt{-g_{MN}+(2\pi\alpha'){F}_{MN}+B_{MN}}\\
        &\supset 
        % -T_8 N_f\left(\frac{8\pi^2}{3}\right)\frac{1}{g_s}(U_\kk R)^{3/2}\int\d^4x\d re^{2\lambda+4\nu+\varphi-\phi}\frac{1}{2}B_{\mu r}^2\\
        % &=-T_8 N_f\left(\frac{8\pi^2}{3}\right)\frac{1}{g_s}(U_\kk R)^{3/2}\int\d^4x\d re^{2\lambda-2\Tilde{\lambda}-4\nu+3\phi-\varphi}\frac{1}{2}\frac{9}{4\Box^2}(-1)\left(\partial_\kappa C_{\rho\sigma}\right)^2\josef{\text{unnötiger Schritt}}\\
        % &=T_8 N_f\left(\frac{8\pi^2}{3}\right)\frac{1}{g_s}(U_\kk R)^{3/2}\int\d^4x\d re^{2\lambda-2\Tilde{\lambda}-4\nu+3\phi-\varphi}\frac{9}{8\Box}(-1)\josef{2}\left(\tilde{C}_{\mu\nu}a(r)\right)^2\\
        %&=
        %-T_8 N_f\left(\frac{8\pi^2}{3}\right)\frac{1}{g_s}(U_\kk R)^{3/2}\int\d^4x\d re^{2\lambda-2\Tilde{\lambda}-4\nu+3\phi-\varphi}\frac{9}{2\Box}a(r)^2\eta^{\mu\nu}V_\mu V_\nu\\
        %&=
        -T_8 N_f\left(\frac{8\pi^2}{3}\right)\int\d^4x\d z\sqrt{-g_{D8}}e^{-\phi}\frac{1}{2}g^{\mu\nu}g^{zz}B_{\mu z}B_{\nu z}\\
        &=
        -\frac{2\lambda^3 N_f N_c}{27 (2\pi)^5 R^6}\int\d^4x\d z(1+z^2)M_4(z)^2\frac{1}{2\Box}\eta^{\mu\nu}V_\mu V_\nu\\
        &=-\int\d^4x\frac{1}{2}\delta \lambda_VM_\kk^2 \eta^{\mu\nu}V_\mu V_\nu,\\  \delta \lambda_V&=\frac{2\lambda^3 N_f N_c}{27 (2\pi)^5 M_\kk^2  R^6 M_V^2}\int\d z(1+z^2)M_4(z)^2=0.00233\lambda^2 \frac{N_f}{N_c}
    \end{split}
    \label{eq:massDBI}
\end{equation}
where we projected out the spin-1 part of $\tilde{C}_{\rho\sigma}(x^\mu)$ with $\tilde{C}_{\rho\sigma}(x^\mu)=\frac{1}{\sqrt{\Box}}\epsilon_{\rho\sigma}^{\ \ \kappa\lambda}\partial_\kappa V_\lambda(x^\mu)$. %The appearance of the eigenvalue in the denominator is due to $1/\sqrt{\Box}$ in the projection. 
Treating this contribution perturbatively we obtain for $N_f=3, N_c=3, \lambda=16.63\dots 12.55$
%$\delta M_V^2=\delta \lambda_V M_\kk^2=0.0215\lambda^2 M_\kk^2\frac{N_f}{N_c}\frac{M_\kk^2}{M_V^2}=0.00233\lambda^2 M_\kk^2\frac{N_f}{N_c}$. For $N_f=3, N_c=3, \lambda=16.63\dots 12.55$ and $M_\kk=949$ MeV this corresponds to 
an increase of the mass of the vector glueball of $100\dots 57$ MeV, i.e., only $3.4\dots 2$\%. 

Since this correction is of the same order as backreaction effects \cite{Burrington:2007qd,Bigazzi:2014qsa} that we
otherwise ignore in the following,\footnote{See Ref.~\cite{Domokos:2022djc} for a recent study of such backreaction effects for the glueballs of the WSS model.}
and since it is numerically quite negligible, we shall later use only
the leading order result for the vector glueball mass.

\subsubsection{Mixing with vector mesons}
\label{sec:vectormixing}

A parametrically more important term of order $\sqrt{N_f/N_c}$ is given by a 
%The flavor branes also introduce a non-diagonal term between 
bilinear term involving
the vector glueball and the singlet flavor gauge field $\hat v=v^{a=0}$. 
Explicitly it is given by
\begin{equation}
    \begin{split}
        S_{DBI}&=-T_8\tr\int\d^9x e^{-\phi}\sqrt{-g_{MN}+(2\pi\alpha'){F}_{MN}+B_{MN}}\\
        &\supset 
        -\int\d^4x \xi_n\eta^{\mu\nu}\hat v_\mu^{n}(x^\mu) V_\nu(x^\mu),\\ \xi_n&=\frac{\kappa\lambda}{2\pi M_V}\frac{M_\kk}{R^3} \tr T^0\int \d z (1+z^2) M_4(z)\psi_{2n-1}'(z) \\
%        &\xi_1=-\frac{\{0.077, 0.071, -0.022\}}{M_V}\tr T^0 M_\kk^3 \frac{\lambda}{\sqrt{N_c}}%=-\{0.025,0.023,-0.007\} M_\kk^2\frac{\lambda\sqrt{N_f/2}}{\sqrt{N_c}}\\
        %&\quad 
        &=\{-0.0180,-0.0165,0.005,\ldots\} \lambda M_\kk^2\sqrt{\frac{N_f}{N_c}}
        % \\
        % &=\{-0.055, -0.050,0.016\} \frac{M_\kk}{M_V} \lambda M_\kk^2\sqrt{\frac{N_f}{N_c}}
    \end{split}
    \label{eq:massMixing}
\end{equation}
for the first three vector meson modes. Note that, as explained above, the integral over $z$ involves $M_4$ as an odd function.
% \hl{F: Last value is correct for $M_V$ arbitrary, the line before used $M_V$ without DBI correction}The integrand diverges as $z\to\infty$ but the boundary condition $M_4(\infty)=0$, enforces a finite result \hl{dann divergiert der Integrand also nicht?! Haengt ja davon ab, wie schnell $M_4$ gegen null geht! F: Man muss hier aufpassen mit welcher WorkingPrecision man die DGL für $M_4$ löst, damit die BCs die Konvergenz sichern. Ungeschickt formuliert, kann man eigentlich weglassen}. Thus, care must be taken in the numerical evaluation of the overlap intergrals. The rescaling of $a(r)$ above corresponds to $a(z)\to z M_4(z)$, which is the same rescaling for $\psi$ as in the original WSS paper. Hence we restriced the integration region to positive $z$ and multiply by a factor 2 when using $M_4$ to make the $z$ parity more pronounced. This convention is implicitly understood in the following. 
Restricting to the ground-state singlet vector meson,
the combined kinetic terms for singlet vector mesons and the vector glueball are then given by
\begin{equation}
    \mathcal{L}^{(2)}_{V,\hat v}=-\int\d^4 x\left(\frac{1}{4}\hat f_{\mu\nu}^2+\frac{1}{2}m^2 \eta^{\mu\nu}\hat v_\mu \hat v_\nu +\xi_1\eta^{\mu\nu} \hat v_\mu V_\nu+\frac{1}{4}F_{\mu\nu}^2+\frac{1}{2}M_V^2 \eta^{\mu\nu}V_\mu V_\nu\right).
\end{equation}
%Since the mixing in Eq.\eqref{eq:massMixing} is a mass mixing, 
With degenerate vector meson masses,
the Lagrangian is readily diagonalized by a unitary field redefinition
\begin{equation}
    \begin{split}
        V_\mu & \to \tilde V_\mu \cos\theta - \tilde v_\mu \sin\theta\\
        \hat v_\mu & \to \tilde V_\mu \sin\theta + \tilde v_\mu\cos\theta 
    \end{split}
\end{equation}
with mixing angle
\begin{equation}
    \theta=\frac{1}{2}\arctan\frac{2\xi_1}{M_V^2-m^2}
    \label{eq:mixingAngle}
\end{equation}
and masses
\begin{equation}
    \begin{split}       \Tilde{m}^2&=m^2\left(\cos^2\theta+\frac{M_V^2}{m^2}\sin^2\theta-\frac{2\xi_1}{m^2}\sin\theta\cos\theta\right),%=(0.658876\dots 0.663367) M_\kk^2,
        \\     \Tilde{M}_V^2&=M_V^2\left(\cos^2\theta+\frac{m^2}{M_V^2}\sin^2\theta+\frac{2\xi_1}{M_V^2}\sin\theta\cos\theta\right).%=(9.23765\dots 9.23316) M_\kk^ 2.
    \end{split}   
\end{equation}

For example, for $N_f=2$, where $\rho$ and $\omega$ are approximately degenerate, we obtain
\begin{equation}
    \theta=-(1.52\dots 1.18)^\circ
\end{equation}
with $M_V=\sqrt{\lambda_V+\delta \lambda_V}M_\kk=(2949\dots2921)$ MeV. After the diagonalization, the masses are only slightly changed and given by
\begin{equation}
    \begin{split}
        \Tilde{m}&=773\dots774\ \mathrm{MeV}\\
        \Tilde{M}_V&=2950\dots2921\ \mathrm{MeV},
    \end{split}
\end{equation}
which would make the $\omega$ meson 2-3 MeV lighter than the $\rho$, while in reality it is roughly 12 MeV heavier.

Larger effects could however arise for vector mesons that are comparable in mass with the vector glueball, such as charmonia, but for those
the WSS model does not provide a reasonable description, because their masses are dominated by the quark masses whereas the vector mesons in the WSS model are independent of quark masses. Nevertheless, we can study the additional decay modes of vector charmonia
that would be contributed by a certain mixing with vector glueballs. We shall return to this question after having determined the decay modes and partial widths of vector glueballs.

\subsection{Decays of the vector glueball}

% When brane actions are considered, the democratic formulation of supergravity \cite{Bergshoeff:2001pv,Tomasiello:2022dwe}, where all field strengths have kinetic terms, the duality relations are imposed, and no (explicit) Chern Simons term appears, must not be used. In this case one thus truncates the kinetic terms at $F_4$ for type IIA, but the dualization relations still hold. However, they give no additional dynamical fields. This is called the \textit{dual formulation of supergravity} of \cite{Bergshoeff:2001pv} where the dualized potentials appear as Lagrange multipliers for the $F_{p+1}$, enforcing the Bianchi identities. For $p$ branes with $p\geq 4$ one needs to use this action and hence the Chern-Simons term sums over all $p$ form degrees.

Except for the mixing term \eqref{eq:massMixing},
all leading-order couplings of the vector glueball with ordinary mesons 
originating from the DBI action vanish, since they involve a trace of commutator terms.
% The leading couplings that would originate from the DBI action through the non-abelian field strengths vanishes, since we trace over commutator terms (see Eq.\eqref{eq:massMixing}). 
Hence to this order all couplings arise through the Chern-Simons term and are thus anomalous. Further we note that $C_3$ is dual to $C_5$
since $F_6=\star F_4$,
%: $\d C_5=F_6=\star F_4=\star \d C_3$ 
leading to contributions from $B_2$ as well as $C_3$.

From the Chern-Simons term of the D8-brane we obtain couplings to mesons, and through VMD also to photons, namely from
\begin{equation}
\begin{split}
S_{CS}^{D8}&=T_8\sum_p\int_{D8}\sqrt{\hat{\mathcal{A}}(\mathcal{R})}\Tr \exp\left(2\pi\alpha' {F}+B\right)\wedge C_p\\
        &\supset T_8\int_{D8}\Tr\frac{(2\pi\alpha')^2}{2!}{F}\wedge{F}\wedge C_5+\Tr\frac{(2\pi\alpha')^2}{2!}{F}\wedge{F}\wedge B_2\wedge C_3.
\end{split}
    \label{eq:chernSimonsContributions}
\end{equation}
Looking at each term separately we have
\begin{eqnarray}
        &&{F}\wedge{F}\wedge C_5%\supset F\wedge{F}\wedge C_5
        =A\wedge{F}\wedge\d C_5=A\wedge{F}\wedge\star \d C_3\label{FFC5}\\
        &&{F}\wedge{F}\wedge B_2\wedge C_3%\supset F\wedge{F}\wedge B_2\wedge C_3
        =A\wedge{F}\wedge B_2\wedge F_4.\label{FFBC}
\end{eqnarray}
In the first term we can use the Hodge dual to fill the indices pertaining to the $S_4$. In the second term we can distribute the indices to obtain the $F_4$ field strength from the background and $B_{\mu z}$. Note that for the field strengths with $p>4$ we have the twisted field strengths \cite{Tomasiello:2022dwe}
\begin{equation}
    F_{p+1}=\d C_p-H\wedge C_{p-2}=(-1)^{p(p-1)/2}\star F_{9-p},\quad p>4
\end{equation}
but they are not dynamical \cite{Bergshoeff:2001pv}. 
%Since the background does not source $B_2$ and no RR fields with even form degree exist in type IIA we only need to be careful with the sign. 

From %\st{the first two lines}
\eqref{FFC5}
we obtain 
\begin{equation}
     \begin{split}
     A\wedge{F}\wedge\star \d C_3 &=-\frac{1}{2}\frac{1}{6 g_s}  \sqrt{{-g}} g^{\tau\tau} \bigg( 
       \left( g^{\mu\sigma}g^{\nu\kappa}g^{\lambda\rho}+2 g^{\mu\kappa}g^{\nu\lambda}g^{\rho\sigma}\right)A_\mu {F}_{\nu\rho}\partial_\sigma C_{\kappa\lambda\tau}
      \\   
     & + g^{zz}g^{\mu\rho}g^{\nu\sigma}\left(A_z{F}_{\mu\nu}+2A_\mu {F}_{\nu z} \right)\partial_z C_{\rho\sigma\tau}\bigg)\d^4x\d z\d \Omega_4\\
     &=-\frac{1}{2}\frac{1}{6g_s}\sqrt{-g}g^{\tau\tau}\bigg(
     a(z)g^{\mu\kappa}g^{\nu\lambda}g^{\rho\delta}A_\mu F_{\nu\rho}\frac{1}{\sqrt{\Box}}\left(\partial_\kappa\star F^V_{\lambda\delta}+\partial_\delta\star F^V_{\kappa\lambda}+\partial_\lambda\star F^V_{\delta\kappa}\right)\\
     +g^{zz}g^{\mu\kappa}g^{\nu\lambda}&\left(2A_z\partial_\mu A_\nu+2A_\mu\partial_\nu A_z-2A_\mu\partial_z A_\nu-3iA_z[A_\mu,A_\nu]\right)\partial_z \frac{a(z)}{\sqrt{\Box}}\star F^V_{\kappa\lambda}\bigg)\d^4x\d z\d \Omega_4\\
     &=-\frac{1}{6g_s}\sqrt{-g}g^{\tau\tau}\bigg(
     -\frac{a(z)}{2}g^{\mu\kappa}g^{\nu\lambda}g^{\rho\delta}A_\mu F_{\nu\rho}\epsilon_{\kappa\lambda\delta\sigma}\sqrt{\Box}\eta^{\sigma\alpha}V_\alpha\\
     +g^{zz}g^{\mu\kappa}g^{\nu\lambda}&\left(A_z\partial_\mu A_\nu+A_\mu\partial_\nu A_z-A_\mu\partial_z A_\nu-\frac{3i}{2}A_z[A_\mu,A_\nu]\right)\partial_z \frac{a(z)}{\sqrt{\Box}}\star F^V_{\kappa\lambda}\bigg)\d^4x\d z\d \Omega_4\\
     % &{\color{blue}\supset}%=
     % -\frac{1}{4}\frac{1}{6g_s}\omega_\tau\sqrt{-g}g^{\tau\tau}\bigg(\\
     % &a(z)g^{\mu\kappa}g^{\nu\lambda}g^{\rho\delta}A_\mu F_{\nu\rho}\frac{1}{\sqrt{\Box}}\left(\partial_\kappa\star F^V_{\lambda\delta}+\partial_\delta\star F^V_{\kappa\lambda}+\partial_\lambda\star F^V_{\delta\kappa}\right)\\
     % &+g^{zz}g^{\mu\kappa}g^{\nu\lambda}\left(2A_z\partial_\mu A_\nu+2A_\mu\partial_\nu A_z-2A_\mu\partial_z A_\nu\right)\partial_z \frac{a(z)}{\sqrt{\Box}}\star F^V_{\kappa\lambda}\bigg)\d^4x\d z\d \Omega_4\\
     % &=-\frac{1}{12g_s}\omega_\tau\sqrt{-g}g^{\tau\tau}\bigg(-\frac{a(z)}{2}g^{\mu\kappa}g^{\nu\lambda}g^{\rho\delta}A_\mu F_{\nu\rho}\epsilon_{\kappa\lambda\delta\sigma}\sqrt{\Box} V^\sigma\\
     % %\left(\partial_\kappa\star F^V_{\lambda\delta}+\partial_\delta\star F^V_{\kappa\lambda}+\partial_\lambda\star F^V_{\delta\kappa}\right)\\
     % &+g^{zz}g^{\mu\kappa}g^{\nu\lambda}\left(A_z\partial_\mu A_\nu+A_\mu\partial_\nu A_z-A_\mu\partial_z A_\nu\right)\partial_z \frac{a(z)}{\sqrt{\Box}}\star F^V_{\kappa\lambda}\bigg)\d^4x\d z\d \Omega_4,\\
    \end{split}   
    \label{eq:LagGVC5}
\end{equation}
% \begin{equation}
%     \begin{split}
%      \left(\frac{3R^3}{g_s}\sqrt{\hat{g}_{S_4}}\right)^{-1}A\wedge{F}\wedge B_2\wedge F_4&= \frac{1}{2\cdot2}\epsilon^{MNOPQ}A_M F_{NO} B_{PQ}\\ 
%     % &=\frac{1}{2}\epsilon^{\mu\nu\rho\sigma}A_\mu F_{\nu\rho} B_{\sigma r}\\
%     % &=-\frac{b(r)}{2}  \delta^{\mu\nu\rho}_{\alpha \beta \gamma} A_\mu F_{\nu\rho} \partial^\alpha \tilde{C}^{\beta\gamma}\\
%     % &=-\frac{b(r)}{2} (2 A_\mu F_{\nu\rho} \partial_\mu \tilde{C}_{\nu\rho}+4 A_\mu F_{\nu\rho} \partial_\rho \tilde{C}_{\mu\nu})\\
%     % &=-b(r)A_\mu F_{\nu\rho}\left(\partial_\mu \tilde{C}_{\nu\rho}+\partial_\rho \tilde{C}_{\mu\nu}+\partial_\nu\Tilde{C}_{\rho\mu}\right)\\
%     % &=-\frac{b(r)}{M_V}A_\mu F_{\nu\rho}\left(\partial_\mu\star F^V_{\nu\rho}+\partial_\rho\star F^V_{\mu\nu}+\partial_\nu\star F^V_{\rho\mu}\right)\\
%     % &=b(r) M_V\epsilon^{\mu\nu\rho\sigma}A_\mu F_{\nu\rho}V_\sigma\\
%     &=2b(r)M_V\epsilon^{\mu\nu\rho\sigma}A_\mu\partial_\nu A_\rho V_\sigma
%     \end{split}
%     \label{eq:LagGVB2}
% \end{equation}
and from \eqref{FFBC}
\begin{equation}
    \begin{split}
     A\wedge{F}\wedge B_2\wedge F_4&= \frac{1}{2\cdot2}\epsilon^{MNOPQ}A_M F_{NO} B_{PQ}\left(\frac{3R^3}{g_s}\right)\d^4x\d z\d \Omega_4\\ 
    &=\frac{1}{2}\epsilon^{\mu\nu\rho\sigma}A_\mu F_{\nu\rho} B_{\sigma z}\left(\frac{3R^3}{g_s}\right)\d^4x\d z\d \Omega_4\\
    &=\frac{a(z)}{z}\epsilon^{\mu\nu\rho\sigma}A_\mu F_{\nu\rho} \frac1{\sqrt{\Box}}V_\sigma\left(\frac{3R^3}{g_s}\right)\d^4x\d z\d \Omega_4,\\
%    &{\color{blue}\supset} 2\frac{a(z)}{z}\epsilon^{\mu\nu\rho\sigma}A_\mu\partial_\nu A_\rho \frac1{\sqrt{\Box}}V_\sigma\left(\frac{3R^3}{g_s}\right)\d^4x\d z\d \Omega_4,\\
    \end{split}
    \label{eq:LagGVB2}
\end{equation}
where %we introduced $\omega_\tau=(\delta(\tau)-\delta(\tau-\pi))\d \tau$ to extend the integration region over the whole spacetime and 
$\star F^V_{\mu\nu}=\sqrt{\Box}\tilde{C}_{\mu\nu}$. 
%Note that since $-\infty<z<\infty$ we also introduced a factor $1/2$ together with $\omega_\tau$ in \eqref{eq:LagGVC5} in order not to overcount.
% To project out the spin 1 part of $\tilde{C}_{\rho\sigma}(x^\mu)$ we used $\tilde{C}_{\rho\sigma}=\frac{1}{\sqrt{\Box}}\epsilon_{\rho\sigma}^{\ \ \kappa\lambda}\partial_\kappa V_\lambda$ and $\star F^V_{\mu\nu}=\frac{1}{2!}\epsilon_{\mu\nu}^{\quad\rho\sigma}F^V_{\rho\sigma}=\epsilon_{\mu\nu}^{\quad\rho\sigma}\partial_\rho V_\sigma=\sqrt{\Box}\tilde{C}_{\mu\nu}$. 
Furthermore we utilized the full antisymmetry to rewrite
\begin{equation}
    \begin{split}
        &\left(\partial_\mu\star F^V_{\nu\rho}+\partial_\nu\star F^V_{\rho\mu}+\partial_\rho\star F^V_{\mu\nu}\right)=-\frac{1}{2}\epsilon_{\mu\nu\rho\sigma}\epsilon^{\sigma\alpha\beta\gamma}\partial_\alpha\star F^V_{\beta\gamma}\\
        &=-\frac{1}{4}\epsilon_{\mu\nu\rho\sigma}\epsilon^{\sigma\alpha\beta\gamma}\partial_\alpha\epsilon_{\beta\gamma\lambda\kappa}F_V^{\lambda\kappa}=-\epsilon_{\mu\nu\rho\sigma}\partial_\alpha F_V^{\alpha\sigma}=-\epsilon_{\mu\nu\rho\sigma}\Box V^\sigma .
    \end{split}
\end{equation}
Interactions between the vector glueball, pseudoscalar mesons, and vector mesons are thus given by
\begin{equation}
    \mathcal{L}_{G_V\Pi v}=-\frac{1}{M_V}g_1^m\tr\left(\Pi \partial_\mu v_\nu^{(m)}+v_\mu^{(m)}\partial_\nu \Pi\right)\star F_{\mu\nu}^{V}
    \label{eq:LagGVsv}
\end{equation}
where
\begin{equation}
    \begin{split}
        g_1^m
        % &=T_8\frac{(2\pi\alpha')^2}{2!g_s}\frac{1}{\sqrt{\kappa\pi}M_\kk}\left(\frac{8\pi^2}{3}\right)\frac{R^3}{6}\int\frac{3}{2}2\d z\frac{1}{z}\psi_{2m-1}(z)\partial_z(z M_4(z))\\
        &=\frac{9}{16}\sqrt{\frac{\kappa}{\pi}}\frac{1}{M_\kk^2 R^3}\int\d z\frac{1}{z}\psi_{2m-1}(z)\partial_z(z M_4(z))=\frac{\{15.04, \dots\}}{\sqrt{\lambda}N_c},
    \end{split}
\end{equation}
and we explicitly pulled out the mass dependence in the Lagrangian and used 
%$A_r\d r=A_z \d z$ and 
$A_z=\Pi(x^\mu)K^{-1}/\sqrt{\kappa\pi M_\kk^2}$. The couplings to vector- and axial vector mesons are governed by
\begin{equation}
    \begin{split}
        \mathcal{L}_{G_V\to va}
        =\frac{1}{M_V}f_1^{mn}\epsilon^{\mu\nu\rho\sigma}\tr\left(v^m_\mu\partial_\nu a^n_\rho+a^n_\mu\partial_\nu v^m_\rho\right)V_\sigma+%\frac{1}{M_V}f_2^{mn}\epsilon^{\mu\nu\rho\sigma}\tr\left(v^m_\mu\partial_\nu a^n_\rho-a^n_\mu\partial_\nu v^m_\rho\right)V_\sigma
        \frac{1}{M_V}f_2^{mn}\tr\left(v^m_\mu a^n_\nu\right)\star F_{\mu\nu},
    \end{split}
    \label{eq:LagGVav}
\end{equation}
where
\begin{equation}
    \begin{split}\label{f1mn}
        f_1^{mn}&=\frac{3}{8}\frac{\kappa}{M_\kk R^3}\int\d z\left( \frac{3}{2}(1+z^2)^{-1/3}\frac{M_V^2}{M_\kk^2}+36\right)
        \psi_{2m-1}(z)\psi_{2n}(z)M_4(z)=\frac{\{177.83,\hdots\}M_\kk}{ N_c\sqrt{\lambda}}\\
        f_2^{mn}&=\frac{3}{8}\frac{\kappa}{M_\kk R^3}\int\d z\left(\frac{3}{2}\frac{1+z^2}{z}\right)\left(\psi_{2m-1}\psi_{2n}'-\psi_{2m-1}'\psi_{2n}\right)\partial_z(z M_4(z))=\frac{\{16.60,\hdots\}M_\kk}{N_c\sqrt{\lambda}}.
    \end{split}
\end{equation}
% In order to get more tractable expression for the couplings to two spin one states, we first partially integrate Eq.\eqref{eq:LagGVC5} to get the four gradient acting on the brane gauge fields, then partially integrate to shuffle one of the r derivatives on the glueball mode to utilize the equations of motion. We thus get
% \begin{equation}
%     \mathcal{L}_{CS}^{D8}\supset \frac{1}{M_V} f_1^{mn}\epsilon^{\mu\nu\rho\sigma}\tr\left(v^m_\mu\partial_\nu a^n_\rho+a^n_\mu\partial_\nu v^m_\rho\right)V_\sigma
% \end{equation}
% where
% \begin{equation}
%     \begin{split}
%         f_1^{mn}&=\frac{3}{8}\frac{\kappa}{M_\kk R^3}\int\d z\left(42\psi_{2m-1}(z)\psi_{2n}(z)M_4(z)\right.\\
%         &\left.+\frac{3}{2}\frac{1+z^2}{z}\left(\psi_{2m-1}(z)\psi_{2n}'(z)+\psi_{2n}(z)\psi_{2m-1}'(z)\right)\partial_z\left(z M_4(z))\right)\right)
%         =\frac{\{177.826,58.912,51.79\}M_\kk}{N_c\sqrt{\lambda}}\\
%     \end{split}
%     \label{eq:LagGVav}
% \end{equation}
%\0{If it were not for the $B_2$ contribution one could utilize the equations of motion and reduce this to a single term. Check signs, prefactors...}
Note that %this coupling has a somewhat softer scaling behavior in $M_\kk$ 
since $M_V\propto M_\kk$, \eqref{eq:LagGVav} does not depend explicitly on the compactification scale. 

The leading quartic couplings are obtained from the commutator terms in the non-abelian field strengths ${F}_{MN}=\partial_M A_N-\partial_N A_M-i\left[A_M,A_N\right]$ of the Chern-Simons interactions. To leading order we have
\begin{equation}
    \begin{split}
        \mathcal{L}_{G_V\to\Pi vv}=\frac{i}{M_V}g_1^{mn}\tr\left(\Pi\left[v_\mu^{(m)},v_\nu^{(n)}\right]\right)\star F_{\mu\nu}^V
    \end{split}
    \label{eq:LagGVsvv}
\end{equation}
with
\begin{equation}
    g_1^{mn}=\frac{9}{16}\sqrt{\frac{\kappa}{\pi}}\frac{1}{M_\kk^2 R^3}\int \frac{3}{2}\d z\frac{1}{z}\psi_{2m-1}(z)\psi_{2n-1}(z)\partial_z(z M_4(z))=\frac{\{1061,\hdots\}}{\lambda N_c^{3/2}}.
\end{equation}
% Noteworthy is the absence of couplings to two pseudoscalars at leading order. These decays are hence suppressed by higher powers of $\alpha'$ in line with expectations from suppressed decays into $K\overline{K}$ in connection to the $\rho\pi$-puzzle \cite{Mo:2006cy}. 
Finally, there are interactions with one axial vector meson and two vector mesons. With the masses obtained by the WSS model, these are however at the mass threshold of the vector glueball, and even above the mass threshold of the pseudovector glueball, which is why they will not be considered in the following.

\subsubsection{Hadronic decays}
From Eq.\eqref{eq:LagGVsv} we obtain the squared amplitude for the decay into one pseudoscalar and one vector meson
\begin{equation}
    \left|\mathcal{M}_{G_V\to\Pi v^m}\right|^2=2\left(g_1^m M_V\tr T_\Pi T_v\right)^2\left(1-2\frac{m_\Pi^2+m_v^2}{M_V^2}+\left(\frac{m_\Pi^2-m_v^2}{M_V^2}\right)^2\right)
\end{equation}
with decay rate
\begin{equation}
    \Gamma_{G_V\to\Pi v^m}=\frac{1}{3}\frac{|\mathbf{p}_v|}{8\pi M_V^2}\left|\mathcal{M}_{G_V\to\Pi v^m}\right|^2.
\end{equation}
The resulting decay rates are collected in Table~\ref{tab:hadronicVectorDecays}.
% \footnote{In  \cite{Freund:1975pn} the decays $G_V\to\rho\pi$ and $G_V\to K\Bar{K}^*$ were studied with a glueball mass of around $M_V=1.41$ GeV and resulting decay rates of $47-90$ MeV and $288-90$ keV respectively. Scaling our results down to that mass we obtain $\Gamma_{G_V\to\rho\pi}=6.67\dots 8.84$ MeV and $\Gamma_{G_V\to K\bar{K}^\star}=111\dots 148$ keV, which is surprisingly in the same ballpark.

From Eq.\eqref{eq:LagGVav} we obtain the squared amplitude for the decay into one axial-vector and one vector meson as
\be
\begin{split}
    &\left|\mathcal{M}_{G_V\to a^m v^n}\right|^2=
    \left(\frac{\tr T_aT_v}{m_a m_v M_V}\right)^2\bigg(
   f_1^{mn} f_2^{mn} \left(m_a^2-m_v^2\right)
   \left(-2 M_V^2 \left(m_a^2+m_v^2\right)+10 m_a^2
   m_v^2+m_a^4+m_v^4+M_V^4\right)\\
   &+\frac{(f_1^{mn})^2}{2M_V^2}\bigg(
   M_V^6 \left(m_a^2+m_v^2\right)-2 M_V^4 \left(6 m_a^2
   m_v^2+m_a^4+m_v^4\right)\\
   &+M_V^2 \left(m_a^2+m_v^2\right) \left(14 m_a^2
   m_v^2+m_a^4+m_v^4\right)+4 m_a^2 m_v^2 \left(m_a^2-m_v^2\right){}^2\bigg)\\
   &+\frac{(f_2^{mn})^2}{2}
  \left(M_V^4 \left(m_a^2+m_v^2\right)-2 M_V^2 \left(-4 m_a^2
   m_v^2+m_a^4+m_v^4\right)+\left(m_a^2-m_v^2\right){}^2
   \left(m_a^2+m_v^2\right)\right)
   \bigg).
\end{split}
\ee

\begin{table}
%    \centering{}\bigskip{}
    \begin{tabular}{lc}
    \toprule
     & $\Gamma_{G_{V(2882)}}%(M_{V}=2882\text{MeV})
     ${[}MeV{]} \tabularnewline\colrule
    $G_{V}\rightarrow\rho\pi$ & 34.3\dots 45.4  \tabularnewline
    $G_{V}\rightarrow K^* {K}$ & 37.8\dots50.1\tabularnewline
    $G_{V}\rightarrow\omega\eta$ & 5.78\dots9.80  \tabularnewline
    $G_{V}\rightarrow \phi\eta$ & 3.45\dots2.81  \tabularnewline
    $G_{V}\rightarrow \omega\eta^\prime$ & 3.06\dots2.50 \tabularnewline
    $G_{V}\rightarrow \phi\eta^\prime$ & 3.22\dots5.46  \tabularnewline\colrule
    $G_{V}\rightarrow a_1\rho,\rho\rho\pi$ & 339\dots417 
    \tabularnewline
%    \st{$G_{V}\rightarrow a_1\rho$} & 822\dots1089  \tabularnewline
%    $G_{V}\rightarrow K_1K^*$ & 625\dots828  \tabularnewline
    $G_{V}\rightarrow K_1(1270)K^*$ & 185\dots246  \tabularnewline
    $G_{V}\rightarrow K_1(1400)K^*$ & 320\dots424  \tabularnewline
    $G_{V}\rightarrow f_1\omega$ & 212\dots281  \tabularnewline
    $G_{V}\rightarrow f_1'\omega$ & 22.4\dots29.7 \tabularnewline
    $G_{V}\rightarrow f_1\phi$ & 9.51\dots12.6  \tabularnewline
    $G_{V}\rightarrow f_1'\phi$ & 47.8\dots63.3  \tabularnewline\colrule
%    \st{$G_{V}\rightarrow \rho\rho\pi$} & 83.8\dots147 \tabularnewline
    $G_{V}\rightarrow K^*K^*\pi$ & 22.7\dots39.9   \tabularnewline
    $G_{V}\rightarrow K^*\rho K$ & 30.3\dots53.2  \tabularnewline
    $G_{V}\rightarrow K^*\omega K$ & 9.85\dots17.3  \tabularnewline
    $G_{V}\rightarrow K^*K^*\eta$ & 7.77\dots12.1 \tabularnewline
    $G_{V}\rightarrow \phi K^* K$ & 3.87\dots6.80  \tabularnewline\colrule 
    $G_{V}\rightarrow \mathrm{hadrons} $& 1301\dots1725\tabularnewline\botrule
    \end{tabular}\caption{Hadronic decays of the vector glueball with WSS model mass $M_{V}=2882\text{ MeV}$ (mixing between vector glueball and singlet vector mesons neglected). Because of the large width of $a_1\to\rho\pi$, the strongly interfering direct and resonant decays into $\rho\rho\pi$ have been combined. 
    %$\theta_V=\arctan 1/\sqrt{2},\ \theta_f=20.4^\circ$.
    }
    \label{tab:hadronicVectorDecays}
\end{table}

For the three-body decays \eqref{eq:LagGVsvv} yields
% Summing over the polarizations of the initial and final states we obtain the squared amplitude to be 
\be
\begin{split}
    &\left|\mathcal{M}_{G_V\to\Pi v^m v^n}\right|^2=\frac{(g_1^{mn})^2}{m_{v_1}^2 m_{v_2}^2
   M_V^2} \bigg(m_{\Pi }^2 \big(m_{v_1}^2
   \left(m_{v_2}^2+M_V^2-s_{12}\right)+M_V^2
   \left(m_{v_2}^2-M_V^2+s_{12}\right)\\
   &+s_{23}
   \left(M_V^2-m_{v_2}^2+s_{12}\right)\big)+m_{v_2}^2 M_V^2 \left(2
   M_V^2-s_{12}\right)
   +s_{23} \left(m_{v_2}^2 \left(s_{12}-4 M_V^2\right)+s_{12}
   \left(M_V^2-s_{12}\right)\right)\\
   &+m_{v_1}^2 \left(m_{v_2}^2 \left(17 M_V^2-3
   \left(s_{12}+s_{23}\right)\right)+m_{v_2}^4+\left(M_V^2-s_{12}\right) \left(2
   M_V^2-2 s_{12}-s_{23}\right)\right)\\
   &+m_{\Pi }^4 \left(-M_V^2\right)+s_{23}^2
   \left(2 m_{v_2}^2-s_{12}\right)+m_{v_1}^4 m_{v_2}^2\bigg)(\mathrm{tr} T_\Pi[T_{v_1}, T_{v_2}])^2,
\end{split}
\ee
where $s_{ij}$ is the center of mass energy of the vector meson and pseudoscalar subsystem. 

\begin{figure}[!htb]
\centering
     \centerline{\includegraphics[width=.2\textwidth,keepaspectratio]{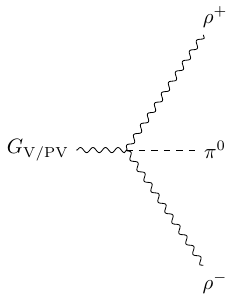}\qquad
     \includegraphics[width=.27\textwidth,keepaspectratio]{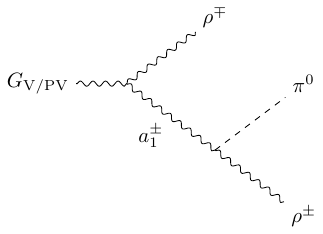}}
     \centerline{\includegraphics[width=.2\textwidth,keepaspectratio]{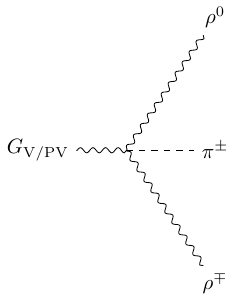}\qquad
     \includegraphics[width=.27\textwidth,keepaspectratio]{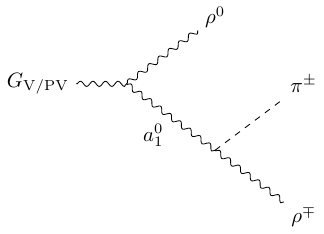}}
    \caption{Feynman diagrams contributing to the hadronic three body decay of the vector glueball into $\rho\rho\pi$}
    \label{fig:FeynVtoPiRhoRho}
\label{fig:had3body}
\end{figure}

Because $a_1$ decays into $\rho\pi$ with a large decay width, which as mentioned above is in fact rather well reproduced by the WSS model, we should consider the decay channels $a_1\rho$ and
$\rho\rho\pi$ together (see Fig.\ \ref{fig:had3body}), since these decays can interfere either positively or negatively.
In fact, we find that there is almost maximal negative interference. In isolation, $G_V\to a_1\rho$ would have a partial width of 822\dots1089 MeV, whereas the resonant
decay $G_V\to a_1\rho\to \rho\rho\pi$ together 
with the nonresonant $G_V\to\rho\rho\pi$ is only about 60\% of that. 

When extending these results to the axial vector mesons involving strange quarks, we instead treat those as narrow resonances and final decay products, neglecting the corresponding interference effects. In fact, in real QCD
the axial vector mesons $K_1$ and $f_1$ have much smaller decay widths. Using their experimental widths indeed leads to comparatively minor changes of the combined resonant plus non-resonant three-body decays.

% In the evaluation of three body decays we did not take into account the possible resonance contributions as depicted in Fig.\ref{fig:had3body} but chose to explicitly state the large decay rates into vector and axial vector mesons in Tab.~\ref{tab:hadronicVectorDecays}. %Overall the resonance contributions turn out to be quite important, especially for the radiative decays, where the resonances fully lie in the integration region. A further intricacy are the $K_1$ mesons, the axial partners of the $K^*$. In the WSS model only the $1^{--}$ and $1^{++}$ spin-1 nonets are present. However the physical states $K_1(1270)$ and $K_1(1400)$ are mixtures of the $1^{++}$ and $1^{+-}$ nonets \PDG.

\subsubsection{Comparison with Ref.~\cite{Giacosa:2016hrm}} %Giacosa et al.}

In Ref.~\cite{Giacosa:2016hrm}, Giacosa et al.\ have
calculated branching ratios for the vector glueball
resulting from three candidate interaction terms in
a chiral Lagrangian inspired by the extended linear sigma model (eLSM) developed in \cite{Parganlija:2012fy,Janowski:2014ppa,Eshraim:2012jv,Eshraim:2016mds}.
Since there is no experimental information on the
coupling constants in either of those terms, ratios of partial decay widths within each of the three possibilities
have been worked out.
Two of these terms involve dimension-4 operators and do not
have a counterpart in the WSS model studied here, so the latter suggests that they may be subleading. A third one breaks dilatation invariance and involves the Levi-Civita tensor that appears also in all the interactions following from the Chern-Simons term in the WSS model, but the resulting interactions differ qualitatively from those considered in
Ref.~\cite{Giacosa:2016hrm}. In particular, there are terms in (\ref{eq:LagGVav}) which
cannot be written in terms of the (dual) field strength tensor for the vector
glueball field, whereas Ref.~\cite{Giacosa:2016hrm} considered only one term
proportional to $\star F^V$.

In Table \ref{tab:hadronicVectorDecaysVsGSJ}, our results for the ratios of the various partial decay widths and $\Gamma(G_V\to\rho\pi)$
are compared with Ref.~\cite{Giacosa:2016hrm}. In both models the dominant decay
mode is $G_V\to a_1\rho$, but in the WSS model this is a factor of 24 larger than
$\Gamma(G_V\to\rho\pi)$, while in the model of Ref.~\cite{Giacosa:2016hrm} this factor is 1.8, more than an order of magnitude smaller.\footnote{Here we are taking into account the substantial negative interference with nonresonant $G_V\to\rho\rho\pi$ decays in the WSS model, while \cite{Giacosa:2016hrm}
considered only two-body decays.}
The second strongest decay mode is $K_1 K^*$, for which Ref.~\cite{Giacosa:2016hrm}
does not list a result, followed by $f_1\omega$.
The WSS model thus predicts a rather strong enhancement of decays
into a pair of axial vector and vector compared to a pair of pseudoscalar and vector.

\begin{table}
    \centering{}\bigskip{}
    \begin{tabular}{lccc}
    \toprule
     & $\frac{\Gamma_{G_{V(2882)}\to\ldots}}{\Gamma_{G_{V(2882)}\to\rho\pi}}$
     & $\frac{\Gamma_{G_{V(3830)}\to\ldots}}{\Gamma_{G_{V(3830)}\to\rho\pi}}$
     & Ref.\cite{Giacosa:2016hrm} \tabularnewline\colrule
         $\rho\pi$ & 1 & 1 & 1  \tabularnewline
    $ K^* {K}$ & 1.1 & 1.21 & 1.3 \tabularnewline
    $\omega\eta$ & 0.17\dots0.22 & 0.18\dots0.23 & 0.16\tabularnewline
    $ \phi\eta$ & 0.10\dots 0.062 &  0.12\dots0.07 & 0.21\tabularnewline
    $ \omega\eta^\prime$ & 0.089\dots 0.055 & 0.11\dots0.07 & 0.13 \tabularnewline
    $ \phi\eta^\prime$ & 0.094\dots 0.12 &  0.14\dots0.18 & 0.18\tabularnewline\colrule
    {$a_1\rho,\ \rho\rho\pi$} &  9.88\dots9.18 & 17.0\dots15.3& {1.8}  \tabularnewline
    $K_1(1270)K^*$ & 5.40 & 12.0 & \tabularnewline
    $K_1(1400)K^*$ & 9.32 & 23.8 &\tabularnewline
    $ f_1\omega$ & 6.2 & 11.8 & 0.55  \tabularnewline
    $ f_1'\omega$ & 0.65 & 1.41 & 0.82 \tabularnewline
    $ f_1\phi$ & 0.28 & 0.83 &\tabularnewline
    $ f_1'\phi$ & 1.4 & 4.92 &\tabularnewline\colrule
%    \st{$ \rho\rho\pi$} & 2.44\dots3.24 & 6.04\dots8.00 & \tabularnewline
    $ K^*K^*\pi$ & 0.66\dots0.88 & 1.92\dots2.54&  \tabularnewline
    $ K^*\rho K$ & 0.88\dots1.17 &3.48\dots4.62 & \tabularnewline
    $ K^*\omega K$ & 0.29\dots0.38 &1.14\dots4.62 &  \tabularnewline
    $ K^*K^*\eta$ & 0.23\dots0.27 & 1.19\dots1.40 & \tabularnewline
    $ \phi K^* K$ & 0.11\dots0.15 & 0.70\dots0.93 & \tabularnewline\botrule
    \end{tabular}\caption{Relative branching ratios of the hadronic decays of the vector glueball with WSS model mass $M_{V}=2882\text{ MeV}$ and with quenched lattice QCD result \cite{Chen:2005mg} 3830 MeV, the latter for the sake of comparison with Ref.\cite{Giacosa:2016hrm}. 
    %Mixing between vector glueball and singlet vector mesons neglected. Without resonance contributions in three body decays. $\theta_V=\arctan 1/\sqrt{2},\ \theta_F=20.4^\circ$.
    }
    \label{tab:hadronicVectorDecaysVsGSJ}
\end{table}

\subsubsection{Radiative decays}
From Eq.\eqref{eq:LagGVsv} we obtain the coupling to photons by utilizing VMD
\begin{equation}
    \mathcal{L}_{G_V\Pi \V}=\frac{1}{M_V}g_1^\V \tr\left(\Pi \partial_\mu\V_\nu+\V_\mu\partial_\nu \Pi\right)\star F^V_{\mu\nu}
    \label{eq:LagGVsV}
\end{equation}
where
\begin{equation}
        g_1^\V=\frac{9}{16}\sqrt{\frac{\kappa}{\pi}}\frac{1}{M_\kk^2 R^3}\int\d z\frac{1}{z}\partial_z(z M_4(z))=\frac{0.31}{\sqrt{N_c}}.
\end{equation}
Employing VMD in Eq.\eqref{eq:LagGVav} we readily obtain the coupling between the vector glueball, an axial vector meson, and one photon as
\begin{equation}
    \mathcal{L}_{CS}^{D8}\supset \frac{1}{M_V} f_1^{\V n}\epsilon^{\mu\nu\rho\sigma}\tr\left(\V_\mu\partial_\nu a^n_\rho+a^n_\mu\partial_\nu \V_\rho\right)V_\sigma+\frac{1}{M_V} f_2^{\V n}\epsilon^{\mu\nu\rho\sigma} \tr\left(\V_\mu\partial_\nu a^n_\rho-a^n_\mu\partial_\nu \V_\rho\right)V_\sigma,
    \label{eq:LagGVaV}
\end{equation}
where
\begin{equation}
    \begin{split}
        % f_1^{\V n}&=\frac{3}{8}\frac{\kappa}{M_\kk^3 R^3}\int\d z\left(42\frac{M_\kk^2}{M_V^2}\right)
        % \psi_{2n}(z)M_4(z)=\frac{\{0.5988,0.30489,0.0295879\}}{M_\kk\sqrt{N_c}}\\
        % f_2^{\V n}&=\frac{9}{8}\frac{\kappa}{M_\kk^3 R^3}\int\d z\frac{1+z^2}{z}\frac{M_\kk^2}{M_V^2}\left(\psi_n'(z)\right)\partial_z\left(z M_4(z))\right)=\frac{\{0.0776,0.0994,0.0575\}}{M_\kk\sqrt{N_c}}
        % f_1^{\V n}&=\frac{3}{8}\frac{\kappa}{M_\kk R^3}\int\d z\left(42\right)
        % \psi_{2n}(z)M_4(z)=\frac{\{5.53,2.81,0.27\}M_\kk}{\sqrt{N_c}}\\
        % f_2^{\V n}&=\frac{9}{8}\frac{\kappa}{M_\kk R^3}\int\d z\frac{1+z^2}{z}\left(\psi_n'(z)\right)\partial_z\left(z M_4(z))\right)=\frac{\{0.72,0.92,0.53\}M_\kk}{\sqrt{N_c}},
        f_1^{\V n}&=\frac{3}{8}\frac{\kappa}{M_\kk R^3}\int\d z\left( \frac{3}{2}(1+z^2)^{-1/3}\frac{M_V^2}{M_\kk^2}+36\right)
        \psi_{2n}(z)M_4(z)=\frac{\{5.88,\hdots\}M_\kk}{\sqrt{N_c}}\\
        f_2^{\V n}&=\frac{3}{8}\frac{\kappa}{M_\kk R^3}\int\d z\left(\frac{3}{2}\frac{1+z^2}{z}\right)\psi_{2n}'(z)\,\partial_z(z M_4(z))=\frac{\{0.36,\hdots\}M_\kk}{\sqrt{N_c}},
    \end{split}
\end{equation}
and the Lagrangian is again independent of the compactification scale.
The quartic coupling including one photon is obtained in a similar fashion from Eq.\eqref{eq:LagGVsvv}
\begin{equation}
    \begin{split}
        \mathcal{L}_{G_V\to\Pi v\V}=\frac{i}{M_V}g_1^{m\V}2\tr\left(\Pi\left[\V_\mu,v_\nu^{(m)}\right]\right)\star F_{\mu\nu}^V,
    \end{split}
    \label{eq:LagGVsvV}
\end{equation}
where
\begin{equation}
    g_1^{m\V}=\frac{27}{32}\sqrt{\frac{\kappa}{\pi}}\frac{1}{M_\kk^2 R^3}\int\d z\frac{1}{z}\psi_{2m-1}(z)\partial_z(z M_4(z))=\frac{\{22.55,\hdots\}}{\sqrt{\lambda} N_c}.
\end{equation}

%\subsubsection{Decay rates}
From Eq.\eqref{eq:LagGVsv} the squared amplitude for the decay into one pseudoscalar and one photon is obtained as
\begin{equation}
    \left|\mathcal{M}_{G_V\to\Pi \V}\right|^2=2\left(eg_1^\V\tr T_\Pi Q\right)^2M_{V}^2\left(1-\frac{m_\Pi^2}{M_V^2}\right)^2
\end{equation}
with decay rate
\begin{equation}
    \Gamma_{G_V\to\Pi \V}=\frac{1}{3}\frac{|\mathbf{p}_v|}{8\pi M_V^2}\left|\mathcal{M}_{G_V\to\Pi \V}\right|^2.
\end{equation}
From Eq.\eqref{eq:LagGVaV} we obtain the squared amplitude for the decay into one axial vector meson and one photon
\begin{equation}
\begin{split}
    \left|\mathcal{M}_{G_V\to a \V}\right|^2=&
    \frac{\tr T_aQ ^2}{2 m_a^2 M_V^4}\bigg(-2 M_V^2 f_1^{n \mathcal{V}} f_2^{n \mathcal{V}} \left(-2 m_a^2 M_V^2-7
   m_a^4+M_V^4\right)\\
   &+(f_1^{ n \mathcal{V}})^2 \left(9 m_a^4 M_V^2-6 m_a^2 M_V^4+4
   m_a^6+M_V^6\right)+M_V^2 (f_2^{n \mathcal{V}} )^2\left(6 m_a^2
   M_V^2+m_a^4+M_V^4\right)\bigg).
\end{split}
\end{equation}
The squared amplitude for the three-body decays resulting from Eq.\eqref{eq:LagGVsvV} is given by
\be
\begin{split}
    &\left|\mathcal{M}_{G_V\to\Pi v^n\V}\right|^2=\frac{2 (g_1^{m \mathcal{V}})^2}{m_v^2 M_V^2(M_V^2-s_{12})^2} \bigg(2 m_v^2 M_V^2
  \left[(s_{12}-M_V^2)(m_{\Pi }^2-3 M_V^2+2
   s_{12})-s_{23}(M_V^2+s_{12})\right]\\
   &\qquad +(m_{\Pi
   }^2-s_{12}){}^2(M_V^2-s_{12})^2+2 s_{12} s_{23}(m_{\Pi
   }^2-s_{12})(M_V^2-s_{12})+2 m_v^4 M_V^4+s_{23}^2
   (M_V^4+s_{12}^2)\bigg).
\end{split}
\ee
There are no three-body decays with two external photons due to the appearance of the commutator in Eq.\eqref{eq:LagGVsvV}, but there are also decays into one photon, one vector meson, and one axial vector meson determined by
% {\color{blue}\\
% Comment out when sending proofs
% \begin{equation*}
%     \epsilon^{\mu\nu\rho\sigma}A_\mu F_{\nu\rho}\supset -i\epsilon^{\mu\nu\rho\sigma}A_\mu [A_\nu, A_\rho]
% \end{equation*}
% There are three ways to distribute two vector and one axial vector meson (can be checked by plugging in $A_\mu=v_\mu+a_\mu$, writing out commutators with $if^{abc}t^c$, utilize $\tr t^a t^b=\delta^{ab}/2$ and renaming $\epsilon$ indices and adjoint indices. In the end replace $f^{abc}=-2i \tr t^a [t^b,t^c]$.
% \begin{equation*}
%     -i\epsilon^{\mu\nu\rho\sigma}A_\mu [A_\nu, A_\rho]\supset -3i \epsilon^{\mu\nu\rho\sigma}\tr v_\mu[v_\nu,a_\rho]
% \end{equation*}
% This is at the mass threshold, so we ignore it. However, we can replace either vector meson with a photon which gives a factor of 2 (can again be checked by writing everything out explicity and substitute the commutators with $if^{abc}t^c$. 
% }
\begin{equation}\label{LGVavV}
    \mathcal{L}_{G_V\to av\mathcal{V}} = -\frac{3i}{M_V}f_1^{mn}\epsilon^{\mu\nu\rho\sigma}\mathrm{tr}\mathcal{V}_\mu [v_\nu^m,a_\rho^n] V_\sigma
\end{equation}
with the same coupling $f_1^{mn}$ as in \eqref{f1mn} that dominated the hadronic decays.

\begin{figure}[!htb]
\centering
% \feynmandiagram [layered layout, horizontal=a to b] {
% a [particle=\(G_{PV}\)] -- [boson] b -- [boson] v1 [particle=\(\gamma\)],
% b -- [scalar] c -- [scalar] s1 [particle=\(\pi\)],
% b -- [boson] v2 [particle=\(\rho\)],
% };
% \feynmandiagram [layered layout, horizontal=a to b] {
% a [particle=\(G_{\mathrm{V/PV}}\)] -- [boson] b -- [boson] v1 [particle=\(\gamma\)],
% b -- [scalar] s1 [particle=\(\pi^\pm\)],
% b -- [boson] v2 [particle=\(\rho^\mp\)],
% };\hspace{3cm}
% \feynmandiagram [layered layout, horizontal=a to b] {
% a [particle=\(G_{1^{\pm-}}\)] -- [boson] b -- [boson] v1 [particle=\(\gamma\)],
% b -- [boson, edge label'=\(a_1^0\)] c,
% c -- [scalar] s1 [particle=\(\pi^\pm\)],
% c -- [boson] v2 [particle=\(\rho^\mp\)],
% };
% \feynmandiagram [layered layout, horizontal=a to b] {
% a [particle=\(G_{PV}\)] -- [boson] b -- [boson,edge label=\(a_1\)] c -- [boson] v1 [particle=\(\rho\)],
% b -- [boson] v2 [particle=\(\gamma\)],
% c -- [scalar] s1 [particle=\(\pi\)]
% };
     % \begin{subfigure}[b]{0.45\textwidth}
     %     \centering
     %     \includegraphics[width=.7\textwidth,keepaspectratio]{img/4PointRhoPiGamma_figure4.pdf}
     %     \caption{$G_V\to\rho^\pm(p_1)\pi^\mp(p_2)\gamma(p_3)$}
     %     \label{fig:FeynVtoPiRhoGamma1}
     % \end{subfigure}
     % \hfill
     % \begin{subfigure}[b]{0.45\textwidth}
     %     \centering
     %     \includegraphics[width=.95\textwidth,keepaspectratio]{img/3PointRhoPiGamma_figure5.pdf}
     %     \caption{$G_V\to\rho^\pm(p_1)\pi^\mp(p_2)\gamma(p_3)$}
     %     \label{fig:FeynVtoPiRhoGamma2}
     % \end{subfigure}
     \centerline{\includegraphics[width=.2\textwidth,keepaspectratio]{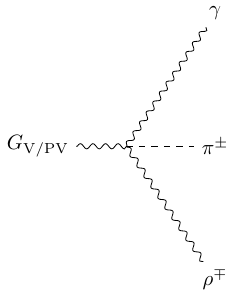}\qquad\includegraphics[width=.27\textwidth,keepaspectratio]{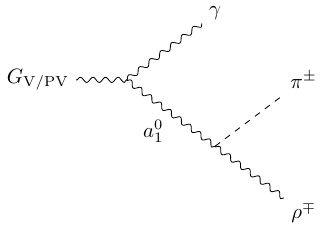}}
\caption{Feynman diagrams contributing to the radiative three body decay of the vector glueball into $\pi\rho\gamma$}
\label{fig:rad3body}
\end{figure}

The various partial decay widths are collected in Table~\ref{tab:radiativeVectorDecays}.
Again we combine $\rho\pi$ decay products with
resonant $a_1\to\rho\pi$ contributions (see Fig.\ \ref{fig:rad3body}), although here the interference is of lesser importance.

% \begin{table}
%     \centering{}\bigskip{}
%     \begin{tabular}{lcc}
%     \toprule
%      & $\Gamma_{G_{V}}(M_{V}=2882\text{MeV})${[}MeV{]} & $\Gamma_{G_{V}}(M_{G}=3850\text{MeV})${[}MeV{]}\tabularnewline\colrule
%     $G_{V}\rightarrow\pi^0\gamma$ & 4.01  & 5.37\tabularnewline
%     $G_{V}\rightarrow\eta\gamma$ & 1.13\dots1.00 & 1.60\dots1.41 \tabularnewline
%     $G_{V}\rightarrow\eta^\prime\gamma$ & 0.06\dots0.16 & 0.09\dots0.25 \tabularnewline\colrule
%     $G_{V}\rightarrow a_1\gamma$ & 0.18 & 0.92 \tabularnewline
%     $G_{V}\rightarrow f_1\gamma$ & 0.03 & 0.18 \tabularnewline
%     $G_{V}\rightarrow f_1'\gamma$ & 0.01 & 0.05 \tabularnewline\colrule
%     $G_{V}\rightarrow \rho\pi\gamma$ & 1.23\dots1.63 &  3.31\dots4.40 \tabularnewline
%     $G_{V}\rightarrow K^*K\gamma$ & 128\dots170 &  1816\dots2401 \tabularnewline\botrule
%     \end{tabular}\caption{Radiative decays of the vector glueball with WSS model mass $M_{V}=2882\text{MeV}$ and extrapolated to the lattice result $M_{V}=3850\text{MeV}$. $\theta_F=20.4^\circ$.}
%     \label{tab:radiativeVectorDecays}
% \end{table}
\begin{table}
    \centering{}\bigskip{}
    \begin{tabular}{lc}
    \toprule
     & $\Gamma_{G_{V(2882)}}%(M_{V}=2882\text{MeV})
     ${[}keV{]} \tabularnewline\colrule
    $G_{V}\rightarrow\pi^0\gamma$ & 27.8  \tabularnewline
    $G_{V}\rightarrow\eta\gamma$ & 7.85\dots 6.96 \tabularnewline
    $G_{V}\rightarrow\eta^\prime\gamma$ & 0.40\dots1.10  \tabularnewline\colrule
    $G_{V}\rightarrow a_1\gamma,\rho\pi\gamma$ & 358\dots361  \tabularnewline
%    \st{$G_{V}\rightarrow a_1\gamma$} & 232  \tabularnewline
    $G_{V}\rightarrow f_1\gamma$ & 41.5 \tabularnewline
    $G_{V}\rightarrow f_1'\gamma$ & 11.4  \tabularnewline\colrule
%    \st{$G_{V}\rightarrow \rho\pi\gamma$} & 146\dots193 \tabularnewline
    $G_{V}\rightarrow K^*K\gamma$ & 78.2\dots104 \tabularnewline\colrule
    $G_V \to a_1 \rho \gamma$          &     338\dots447 \tabularnewline
$G_V \to K_1(1270) K^* \gamma$         &     47.2\dots62.6 \tabularnewline
$G_V \to K_1(1400) K^* \gamma$         &     47.3\dots62.7 \tabularnewline\colrule
    $G_V\to X+\gamma$ & 958\dots1126 \tabularnewline\botrule
    \end{tabular}
    \caption{Radiative decays of the vector glueball with WSS model mass $M_{V}=2882\text{ MeV}$. 
    %No resonance contributions in three body decays. $\theta_F=20.4^\circ$.
    }
    \label{tab:radiativeVectorDecays}
\end{table}

\subsubsection{Implications for the \texorpdfstring{$\rho\pi$}{rho pi} puzzle}

A long-standing puzzle in charmonium physics is the experimental fact that the radial excitation $\psi'=\psi(2S)=\psi(3686)$ of the vector meson $J/\psi$ has decays into $\rho\pi$, $K^* K$, and other hadronic channels with partial widths far below the expectation from their nature of a nonrelativistic bound state of $c$ and $\bar c$ \cite{Chen:1998ma,Mo:2006cy}.

Early attempts to explain this are based on
a mixing of the ground state $J/\psi$ with a vector glueball that enhances the decay modes
involved in the $\rho\pi$ puzzle \cite{Freund:1975pn,Hou:1982kh,Brodsky:1987bb,Chan:1999px,Hou:1996kw,Hou:1997it}, for instance by assuming a narrow vector glueball with mass close to that of $J/\psi$ so that a resonant
enhancement of the mixing appears (cf.\ \eqref{eq:mixingAngle}).

The WSS model is certainly not suitable to describe the nonrelativistic $c\bar c$ bound states, but it makes concrete predictions for the decays of the vector glueball.
Since the vector glueball is predicted to be a rather wide resonance, it does not fit the picture assumed in \cite{Brodsky:1987bb}. Moreover, lattice QCD predicts a mass of the vector glueball about 700 MeV higher than that of $J/\psi$. Nevertheless, it is not excluded that the mixing of the vector glueball could be strongly different for the different vector charmonia. Indeed, in Sec.\ \ref{sec:vectormixing}, we have found that the
mixing of excited vector mesons depends strongly on the mode number, albeit the first two modes happened to be comparable, but that need not be the case for vector mesons far from the chiral limit.

However, the decay pattern that we have obtained for the vector glueball makes it rather unsuitable for an explanation of the $\rho\pi$ puzzle. While the vector glueball has $\rho\pi$ and $K^*K$ as important decay modes, decays into $a_1\rho$ and $K_1(1400)K^*$ are much stronger,
but have not been observed in the hadronic decays of $J/\psi$ \PDG.

% The mixing angle obtained above for $N_f=3$ light flavors points to an almost unmixed vector glueball and $\omega_0\ (\omega\ \mathrm{and}\ \phi)$ meson. It is also  in agreement with the value of $|\theta|<2^\circ$ quoted in \cite{Giacosa:2016hrm} for the resolution of the $\rho\pi$ puzzle. The $\rho\pi$ puzzle is the observed strong suppression of the decay channel $\Psi(2S)\to\rho\pi$ which is considered a  radial excitation of the $J/\psi$ meson. A small mixing of the vector glueball with $J/\psi$ can explain such a suppression \cite{Freund:1975pn,Hou:1982kh,Brodsky:1987bb,Chan:1999px,Hou:1996kw,Hou:1997it}. Taking Eq.\eqref{eq:mixingAngle} as a general estimate, and introducing the $J/\psi$ with generator $T^{J/\psi}=\sqrt{1/2}(0,0,0,1)$) by separating one heavy brane and plugging in the PDG value for $m_{J/\psi}=3090$ MeV and the lattice result for $M_V=3850$ MeV \cite{Morningstar:1999rf} we obtain 
% \begin{equation}
%     \theta=-(1.27\dots 0.96)^\circ.
% \end{equation}
% for $N_c=3, N_f=4$, which is well below the bound of $2^\circ$. 

\section{Revisiting the Pseudovector Glueball}
\label{sec:pvgb}

After Kaluza-Klein reduction of the 3 form field $A_3$ of the 11D supergravity theory to 10D, the $1^{+-}$ glueball is identified with the fluctuations of $B_{\mu\nu}=A_{\mu\nu 11}$ and $C_{\mu \tau r}=A_{\mu \tau r}$. In 10D notation the equations of motion are solved by
% \begin{equation}
%     B_{\mu\nu}=c(r) \tilde{B}_{\mu\nu}(x^\mu),\quad C_{\mu\tau r}=\frac{3}{2g_s\Box }c(r)e^{\tilde{\lambda}-\varphi-4\nu}\eta_{\mu\nu}\epsilon^{\nu\rho\sigma\kappa}\partial_{\rho}\Tilde{B}_{\sigma\kappa}(x^\mu)
% \end{equation}
% \hl{F:new}
\begin{equation}
    B_{\mu\nu}=c(u) \tilde{B}_{\mu\nu}(x^\mu),\quad C_{\mu\tau u}=\frac{3}{2g_s\Box }\frac{c(u)}{u}\eta_{\mu\nu}\epsilon^{\nu\rho\sigma\kappa}\partial_{\rho}\Tilde{B}_{\sigma\kappa}(x^\mu).
\end{equation}
% \begin{equation}
%     B_{\mu\nu}=c(z) \tilde{B}_{\mu\nu}(x^\mu),\quad C_{\mu\tau z}=\frac{1}{g_s\Box }\frac{z}{1+z^2}c(z)\eta_{\mu\nu}\epsilon^{\nu\rho\sigma\kappa}\partial_{\rho}\Tilde{B}_{\sigma\kappa}(x^\mu)
% \end{equation}
Upon rescaling $c(u)=(r/r_\kk)^3N_4(r),\ u^3=r^6/r_\kk^6,$ the radial mode corresponds to the one already obtained in \cite{Brower:2000rp}. In terms of the $z$ coordinate this rescaling amounts to $c(z)=\sqrt{1+z^2}N_4(z)$, hence the $N_4$ mode has even $z$ parity.

In \cite{Brunner:2018wbv} only the Chern-Simons couplings arising from $B_2$ were considered. However, there is an additional coupling arising from the dualization of $F_6=\star F_4$ which has been overlooked. Inducing the pseudovector fluctuation on this term we obtain additionally
\begin{equation}
    \begin{split}
     A\wedge{F}\wedge\star \d C_3 \wedge \omega_\tau&=\frac{1}{4!}  \sqrt{{-g}} A_M F_{NO} F_4^{\tau M N O}\d^4x \d z\d \Omega_4 \\
      &=\frac{1}{3!M_{PV}} \sqrt{{-g}}g^{zz}g^{\tau\tau}  ( A_z F_{\mu \nu}  + 2 A_\mu F_{\nu z})F^{\mu\nu}_{\tilde{V}}\frac{3}{2g_s}z^2c(z)\d^4x \d z\d \Omega_4,\\
    \end{split}
\end{equation}
% \begin{equation}
%     \begin{split}
%      A\wedge{F}\wedge\star \d C_3&=\frac{1}{4!}  \omega_\tau \sqrt{{-g}} A_M F_{NO} F_4^{\tau M N O} \\
%       &=\frac{1}{3!M_{PV}}  \omega_\tau \sqrt{{-g}}g^{zz}g^{\tau\tau}  ( A_z F_{\mu \nu}  + 2 A_\mu F_{\nu z})F^{\mu\nu}_{\tilde{V}}\frac{3}{2g_s}(u^3-1)c(u),\\
%     \end{split}
% \end{equation}
besides 
the couplings already computed in \cite{Brunner:2018wbv}
% \begin{equation}
%     \begin{split}
%      A\wedge{F}\wedge B_2\wedge F_4 
%     %  &=\frac{1}{2\cdot2}\epsilon^{MNOPQ}A_M F_{NO} B_{PQ}\\ 
%     % &=\frac{1}{2\cdot 2}\epsilon^{\mu\nu\rho\sigma}(A_r F_{\mu \nu}+ 2 A_\mu F_{ \nu r} ) B_{\rho \sigma}\\
%     &=-\frac{1}{2}\frac{c(r)}{M_{PV}}(A_r F_{\mu\nu}+2A_\mu F_{\nu r})F_{\mu\nu}^{\Tilde{V}}\left(\frac{3R^3}{g_s}\sqrt{\hat{g}_{S_4}}\right)\d^4x \d r\d \Omega_4.\\
%     % &=\frac{1}{2}\epsilon^{\mu\nu\rho\sigma}\left(A_r\partial_\mu A_\nu+A_\mu F_{\nu r}\right)B_{\rho\sigma}\\
%     % &=\left(A_r\partial_\mu A_\nu+A_\mu F_{\nu r}\right) \Bd^{\mu\nu}
%     \end{split}
% \end{equation}
% \hl{F new}
% \begin{equation}
%     \begin{split}
%      A\wedge{F}\wedge B_2\wedge F_4 
%     &=-\frac{c(u)}{2M_{PV}}(A_u F_{\mu\nu}+2A_\mu F_{\nu u})F_{\mu\nu}^{\Tilde{V}}\left(\frac{3R^3}{g_s}\sqrt{\hat{g}_{S_4}}\right)\d^4x \d u\d \Omega_4.\\
%     \end{split}
% \end{equation}
\begin{equation}
    \begin{split}
     A\wedge{F}\wedge B_2\wedge F_4 
    &=-\frac{c(z)}{2M_{PV}}(A_z F_{\mu\nu}+2A_\mu F_{\nu z})F_{\mu\nu}^{\Tilde{V}}\left(\frac{3R^3}{g_s}\right)\d^4x \d z\d \Omega_4.\\
    \end{split}
\end{equation}
% with $\Bd^{\mu\nu}=\frac{1}{2} \epsilon^{\mu \nu \rho \sigma} \Tilde{B}_{\rho \sigma}$.
From this we obtain
\begin{equation}
    \mathcal{L}_{G_{PV}\to\Pi v}=
    -\left(1-\frac{1}{3!}\right)\frac{1}{M_{PV}}b_1^m \tr \left(v_\mu^{(m)}\partial_\nu\Pi+\Pi \partial_\mu v_\nu ^{(m)}\right) F_{\mu\nu}^{\tilde{V}}
    %+\frac{1}{3!}\frac{1}{M_{PV}}b_1^m \tr \left(v_\mu^{(m)}\partial_\nu\Pi+\Pi \partial_\mu v_\nu ^{(m)}\right) F_{\mu\nu}^{\tilde{V}},
\end{equation}
where the first term is the one already obtained in \cite{Brunner:2018wbv}, and the second term involving $-\frac{1}{3!}$ arises through the dualization of $C_3$,
with
\begin{equation}
    \begin{split}
        b_1^m&=T_8\frac{(2\pi\alpha')^2}{2!}\frac{3R^3}{g_s}\left(\frac{8\pi^2}{3}\right)\int\d z K^{-1/2}\psi_{2m-1}(z)N_4(z)\\
        &=\frac{27}{4}\sqrt{\frac{\kappa}{\pi}}\frac{1}{M_\kk^2R^3}\int\frac{\d z}{\sqrt{1+z^2}}\psi_{2m-1}(z)N_4(z)=\frac{\{112.054,\hdots\}}{\sqrt{\lambda}N_c}.
    \end{split}   
\end{equation}
This results in a reduction of the decay rates of roughly 30\%.

% \begin{figure}[!htb]
%     \centering
%     \includegraphics[width=.65\textwidth,height=.65\textheight,keepaspectratio]{img/PVpi2rhoDecay.pdf}
%     \caption{Decay rate of the Pseudovector glueball in $\rho\rho\pi$ as a function of the $a_1$ decay width. The red shaded region corresponds to the values quoted in the PDG \PDG and the black line to the result obtained in \cite{Brunner:2018wbv} where the resonance and dualized coupling where not taken into account. The green lines correspond to the result without the resonance.}
%     \label{fig:PVpi2rhoDecay}
% \end{figure}
% \begin{figure}[!htb]
%     \centering
%     \includegraphics[width=.65\textwidth,height=.65\textheight,keepaspectratio]{img/PVpirhogammaDecay.pdf}
%     \caption{Decay rate of the Pseudovector glueball in $\pi\rho\gamma$ as a function of the $a_1$ decay width. The red shaded region corresponds to the PDG average. The green band is the result without the resonance.}
%     \label{fig:PVpirhogammaDecay}
% \end{figure}
\begin{table}
    \centering{}\bigskip{}
    \begin{tabular}{lc}
    \toprule
    & $\Gamma_{G_{PV(2311)}}%(M_{G}=2311\text{MeV})
    ${[}MeV{]}\tabularnewline\colrule
    $G_{PV}\rightarrow\rho\pi$ & 585\dots775  \tabularnewline
    $G_{PV}\rightarrow K^* K$ &  259\dots338  \tabularnewline
    $G_{PV}\rightarrow\eta\omega$ &  83.2\dots141  \tabularnewline
    $G_{PV}\rightarrow \eta \phi$ &  13.8\dots11.3  \tabularnewline
    $G_{PV}\rightarrow\eta^\prime \omega$ &  31.9\dots26.0  \tabularnewline
    $G_{PV}\rightarrow \eta^\prime \phi$ & 5.21\dots8.83  \tabularnewline\hline
    $G_{PV}\rightarrow a_1\rho,\rho\rho\pi$ &  433\dots 751  \tabularnewline
    %\st{$G_{PV}\rightarrow a_1\rho$} &  206\dots 273  \tabularnewline
    $G_{PV}\rightarrow K_1(1270) K^*$ &  26.9\dots 35.6  \tabularnewline
    $G_{PV}\rightarrow K_1(1400) K^*$ &  1.72\dots2.82 \tabularnewline
    $G_{PV}\rightarrow f_1\omega$ &  40.9\dots54.2 \tabularnewline
    $G_{PV}\rightarrow f_1'\omega$ &  1.32\dots1.75 \tabularnewline\hline
    %\st{$G_{PV}\rightarrow \rho\rho\pi$} & 428\dots752  \tabularnewline
    $G_{PV}\rightarrow K^*K^*\pi $ & 37.6\dots66.0 \tabularnewline
    $G_{PV}\rightarrow K^*\rho K$ & 5.85\dots10.3    \tabularnewline
    $G_{PV}\rightarrow K^*\omega K$ & 1.66\dots2.91  \tabularnewline\colrule
    $G_{PV}\to $ hadrons & 1476\dots 2162 \tabularnewline\botrule
    \end{tabular}\caption{%\hl{F: $K_1(1400)K^*$ very close to mass threshold. Same for $f_1'\omega$! } 
    Hadronic decays of the pseudovector glueball with WSS model mass of $M_{PV}=2311$ MeV. 
    % $\theta_V=\arctan 1/\sqrt{2}$ and $\theta_K=33^\circ$.
    }
    \label{tab:hadronicPVdecays}
\end{table}

\begin{table}
    \centering{}\bigskip{}
    \begin{tabular}{lcc}
    \toprule
     & $\frac{\Gamma_{G_{PV(2311)}\to\ldots}}{\Gamma_{G_{PV(2311)}\to\rho\pi}}$
     & $\frac{\Gamma_{G_{PV(2980}\to\ldots}}{\Gamma_{G_{PV(2980)}\to\rho\pi}}$
     \tabularnewline\colrule
         $\rho\pi$ & 1 & 1   \tabularnewline
    $ K^* {K}$ & 0.55 &  0.75\tabularnewline
    $\omega\eta$ & 0.14\dots0.18 & 0.17\dots0.21\tabularnewline
    $ \phi\eta$ & 0.02\dots0.01  & 0.04\dots0.03\tabularnewline
    $ \omega\eta^\prime$ & 0.05\dots0.03 & 0.09\dots0.06 \tabularnewline
    $ \phi\eta^\prime$ & 0.009\dots0.01 & 0.04\dots0.05\tabularnewline\colrule
    $ a_1\rho,\rho\rho\pi$ & 0.74\dots0.97 &  2.64\dots 3.35\tabularnewline
%    \st{$ a_1\rho$} & 0.35 & 0.65 \tabularnewline
    $K_1(1270)K^*$ & 0.05  & 0.16  \tabularnewline
    $K_1(1400)K^*$ & 0.003 & 0.24 \tabularnewline
    $ f_1\omega$ & 0.07 & 0.16  \tabularnewline
    $ f_1'\omega$ & 0.002 & 0.015 \tabularnewline
    $ f_1\phi$ & - & 0.01 \tabularnewline
    $ f_1'\phi$ & - & 0.04 \tabularnewline\colrule
%    \st{$ \rho\rho\pi$} & 0.73\dots0.97 & 2.22\dots2.94\tabularnewline
    $ K^*K^*\pi$ & 0.06\dots0.09 & 0.43\dots0.57\tabularnewline
    $ K^*\rho K$ & 0.010\dots0.013 & 0.52\dots0.69\tabularnewline
    $ K^*\omega K$ & 0.003\dots0.004 & 0.17\dots0.22\tabularnewline
    $ K^*K^*\eta$ & - & 0.11\dots0.12\tabularnewline
    $ \phi K^* K$ & - & 0.04\dots0.06 \tabularnewline\botrule
    \end{tabular}\caption{Hadronic decays of the pseudovector glueball with WSS model mass $M_{PV}=2311\text{ MeV}$ and the quenched lattice value of 2980 MeV. 
    % Mixing between vector glueball and singlet vector mesons neglected. Without resonance contributions in three body decays. $\theta_V=\arctan 1/\sqrt{2},\ \theta_F=20.4^\circ$.
    }
    \label{tab:hadronicPVectorDecaysVsGSJ}
\end{table}

\begin{table}
    \centering{}\bigskip{}
    \begin{tabular}{lc}
    \toprule
    & $\Gamma_{G_{PV(2311)}}%(M_{G}=2311\text{MeV})
    ${[}keV{]}\tabularnewline\colrule
    $G_{PV}\rightarrow\pi^0\gamma$ & 0.01\tabularnewline
    $G_{PV}\rightarrow\eta\gamma$ & 1.11\dots0.98 \tabularnewline
    $G_{PV}\rightarrow\eta^\prime\gamma$ & 0.59\dots1.62 \tabularnewline\hline
    $G_{PV}\rightarrow a_1\gamma,\rho\pi\gamma$ & 1395\dots1848   \tabularnewline
%    \st{$G_{PV}\rightarrow a_1\gamma$} & 27.4  \tabularnewline
    $G_{PV}\rightarrow f_1\gamma$ & 5.16  \tabularnewline
    $G_{PV}\rightarrow f_1'\gamma$ & 1.40  \tabularnewline\hline
%    \st{$G_{PV}\rightarrow\rho\pi\gamma$} & 1038\dots 1376  \tabularnewline
    $G_{PV}\rightarrow K^*K\gamma$ & 266\dots 353  \tabularnewline\hline
    $G_{PV}\to X+\gamma $ &1669\dots 2209 \tabularnewline\botrule
    \end{tabular}\caption{Radiative decays of the pseudovector glueball with WSS model mass of $M_{PV}=2311$ MeV. 
    % $\theta_V=\arctan 1/\sqrt{2}$ and $\theta_K=33^\circ$.
    }
    \label{tab:radiativePVdecays}
\end{table}

The corresponding coupling to the photon is readily obtained as
\begin{equation}
    \mathcal{L}_{G_{PV}\to\Pi \V}=
    -\frac{5}{6}\frac{1}{M_{PV}}b_1^\V \tr \left(\V_\mu\partial_\nu\Pi+\Pi \partial_\mu \V_\nu \right) F_{\mu\nu}^{\tilde{V}},
\end{equation}
where
\begin{equation}
    b_1^\V=\frac{27}{4}\sqrt{\frac{\kappa}{\pi}}\frac{1}{M_\kk^2R^3}\int\frac{\d z}{\sqrt{1+z^2}}N_4(z)=\frac{2.70}{\sqrt{N_c}}.
\end{equation}

There is also a coupling between the pseudovector glueball and vector- and axial vector mesons present, which has not been considered in \cite{Brunner:2018wbv}. Their masses are, however, at the threshold of the WSS model mass. Explicitly it is given by
\begin{equation}
    \mathcal{L}_{G_{PV}\to va}=
    -\frac{5}{6}\frac{1}{M_{PV}}b_3^{mn} \tr \left(v_\mu^{(m)}a_\nu ^{(n)}\right) F_{\mu\nu}^{\tilde{V}}
    %+\frac{1}{3!}\frac{1}{M_{PV}}b_3^{mn} \tr \left(v_\mu^{(m)}a_\nu ^{(m)}\right) F_{\mu\nu}^{\tilde{V}},
\end{equation}
with
\begin{equation}
    b_3^{mn}=\frac{27}{4}\frac{\kappa}{M_\kk R^3}\int\d z\sqrt{1+z^2}(\psi_{2m-1}(z)\psi_{2n}'(z)-\psi_{2m-1}'(z)\psi_{2n}(z))N_4(z)=\frac{\{118.66,\hdots\}M_\kk}{\sqrt{\lambda}N_c}.
\end{equation}
This entails a coupling to photons and axial vector mesons given by
\begin{equation}
    \mathcal{L}_{G_{PV}\to \V a}=
    -\frac{5}{6}\frac{1}{M_{PV}}b_3^{m\V} \tr \left(\V_\mu a_\nu ^{(m)}\right) F_{\mu\nu}^{\tilde{V}}
    %+\frac{1}{3!}\frac{1}{M_{PV}}b_3^{m\V} \tr \left(\V_\mu a_\nu ^{(m)}\right) F_{\mu\nu}^{\tilde{V}},
\end{equation}
with
\begin{equation}
    b_3^{m\V}=\frac{27}{4}\frac{\kappa}{M_\kk R^3}\int\d z\sqrt{1+z^2}\psi_{2m-1}(z)\psi_{2n}'(z)N_4(z)=\frac{\{1.75,\hdots\}M_\kk}{\sqrt{N_c}}.
\end{equation}

Three-body decays result from the interactions governed by
\begin{align}
    \mathcal{L}_{G_{PV}\Pi vv}&=\frac{5}{6}\frac{i}{M_V}b_2^{mn}\tr\left(\Pi\left[v_\mu^{(m)},v_\nu^{(n)}\right]\right)F_{\mu\nu}^{\tilde{V}}%-\frac{i}{3!}\frac{1}{M_V}b_2\tr\left(\Pi\left[v_\mu,v_\nu\right]\right)F_{\mu\nu}^{\tilde{V}},
\end{align}
where\footnote{Our results for $b_1^m$ and $b_2^{mn}$ for $m=n=1$ differ from the ones in \cite{Brunner:2018wbv} by factors of 2 and $2^{3/2}$, respectively, due to the different normalization of the $SU(N_f)$ generators.}
\begin{equation}
    b_2^{mn}=\frac{81}{8}\sqrt{\frac{\kappa}{\pi}}\frac{1}{M_\kk^2R^3}\int\d z\psi_{2m-1}(z)\psi_{2n-1}(z)N_4(z)=\frac{\{7257.92,\hdots\}}{\lambda N_c^{3/2}},
\end{equation}
with a corresponding photon coupling given by
\begin{align}
    \mathcal{L}_{G_{PV}\Pi v\V}&=\frac{5}{6}\frac{i}{M_V}2b_2^{m\V} \tr\left(\Pi\left[v_\mu^{(m)},\V_\nu\right]\right)F_{\mu\nu}^{\tilde{V}}
\end{align}
with
\begin{equation}
    b_2^{m\V}=\frac{\{168.081,\hdots\}}{\sqrt{\lambda}N_c}.
\end{equation}
The results for the 
hadronic 
decay rates are collected in Table~\ref{tab:hadronicPVdecays}. 
% %and 
% to be compared with the old results from \cite{Brunner:2018wbv}, %and 
% which are
% extended to include decays of $G_{PV}\to a v$ as well. 
Table \ref{tab:hadronicPVectorDecaysVsGSJ} shows the change of the decay pattern when the WSS model mass $M_{PV}=2311\text{ MeV}$ is replaced by the quenched lattice value of 2980 MeV.
The radiative decays are displayed in Table \ref{tab:radiativePVdecays};
note that there is no analog of \eqref{LGVavV} and thus there are no $av\gamma$ decays.
%Overall the resonance contributions lead to a slight decrease in the decay rates. The dependence of the $G_{PV}\to\rho\rho\pi$ decay rate on the $a_1$ width, which is the largest contributing three body-decay, is plotted in figure \ref{fig:PVpi2rhoDecay} and the radiative decay $G_{PV}\to\pi\rho\gamma$ is plotted in figure \ref{fig:PVpirhogammaDecay}.

%Dalitz-Plots für GV und \hl{GPV}.
% \begin{figure}
% \centering    
% \begin{subfigure}
% \includegraphics{img/dalitzVtoPiRhoRho_s12s13.pdf}
%     \caption{Dalitz plot for $G_V\to\rho^+(p_1)\pi^0(p_2)\rho^-(p_3)$}
%     \label{fig:DalitzVtoPiRhoRho}    
% \end{subfigure}
% \hfill
% \begin{subfigure}   \includegraphics{img/dalitzVtoPipRhoRho_s12s13.pdf}
%     \caption{Dalitz plot for $G_V\to\rho^+(p_1)\pi^0(p_2)\rho^-(p_3)$}
%     \label{fig:DalitzVtoPi+RhoRho}    
% \end{subfigure}
% \caption{Dalitz plots for the decays $G_V$}
% \label{fig:GVDalitz}
% \end{figure}

\begin{figure}
     % \centering
     % \begin{subfigure}[b]{0.45\textwidth}
     %     \centering
     %     \includegraphics[width=\textwidth]{img/dalitzVtoPiRhoRho_s12s13.pdf}
     %     \caption{$G_V\to\rho^+(p_1)\pi^0(p_2)\rho^-(p_3)$}
     %     \label{fig:DalitzVtoPiRhoRho}
     % \end{subfigure}
     % \hfill
     % \begin{subfigure}[b]{0.45\textwidth}
     %     \centering
     %     \includegraphics[width=\textwidth]{img/dalitzVtoPipmRhoRho_s23s13.pdf}
     %     \caption{$G_V\to\rho^0(p_1)\pi^\pm(p_2)\rho^\mp(p_3)$}
     %     \label{fig:DalitzVtoPi+-RhoRho}
     % \end{subfigure}
     \centerline{\includegraphics[width=.42\textwidth]{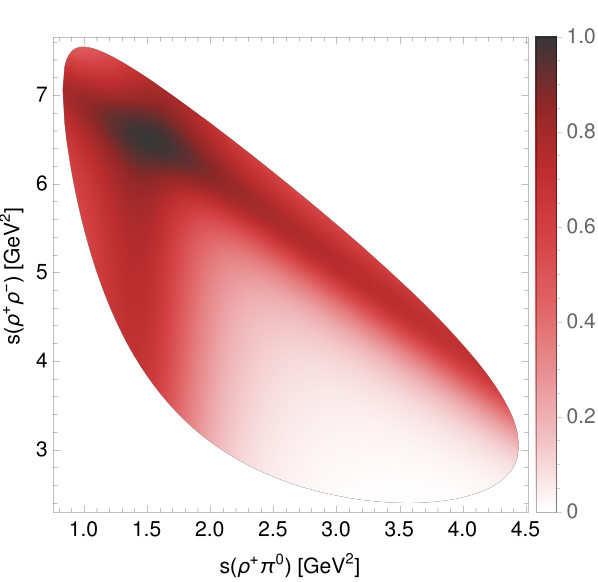}\qquad\includegraphics[width=.42\textwidth]{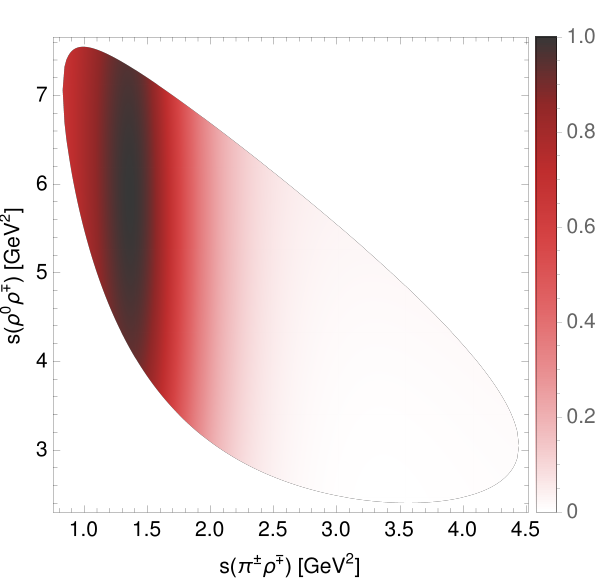}}
        \caption{Dalitz plots for the three body decay $G_V\to\rho\rho\pi$}
        \label{fig:VectorDalitz}
\end{figure}

% \begin{figure}
%      \centering
%      \begin{subfigure}[b]{0.45\textwidth}
%          \centering
%          \includegraphics[width=\textwidth]{img/dalitzVtoPiRhoRho_s12s13.pdf}
%          \caption{$G_V\to\rho^+(p_1)\pi^0(p_2)\rho^-(p_3)$}
%          %\label{fig:DalitzVtoPiRhoRho}
%      \end{subfigure}
%      \hfill
%      \begin{subfigure}[b]{0.45\textwidth}
%          \centering
%          \includegraphics[width=\textwidth]{img/dalitzVtoPimRhoRho_s12s13.pdf}
%          \caption{$G_V\to\rho^0(p_1)\pi^-(p_2)\rho^+(p_3)$}
%          %\label{fig:DalitzVtoPi+RhoRho}
%      \end{subfigure}
%         \caption{Dalitz plots for the three body decay $G_V\to\rho\rho\pi$}
%         %\label{fig:VectorDalitz}
% \end{figure}

\begin{figure}
     % \centering
     % \begin{subfigure}[b]{0.45\textwidth}
     %     \centering
     %     \includegraphics[width=\textwidth]{img/dalitzPVtoPiRhoRho_s12s13.pdf}
     %     \caption{$G_{PV}\to\rho^+(p_1)\pi^0(p_2)\rho^-(p_3)$}
     %     \label{fig:DalitzPVtoPiRhoRho}
     % \end{subfigure}
     % \hfill
     % \begin{subfigure}[b]{0.45\textwidth}
     %     \centering
     %     \includegraphics[width=\textwidth]{img/dalitzPVtoPipmRhoRho_s23s13.pdf}
     %     \caption{$G_{PV}\to\rho^0(p_1)\pi^\pm(p_2)\rho^\mp(p_3)$}
     %     \label{fig:DalitzPVtoPi+-RhoRho}
     % \end{subfigure}
     \centerline{\includegraphics[width=.42\textwidth]{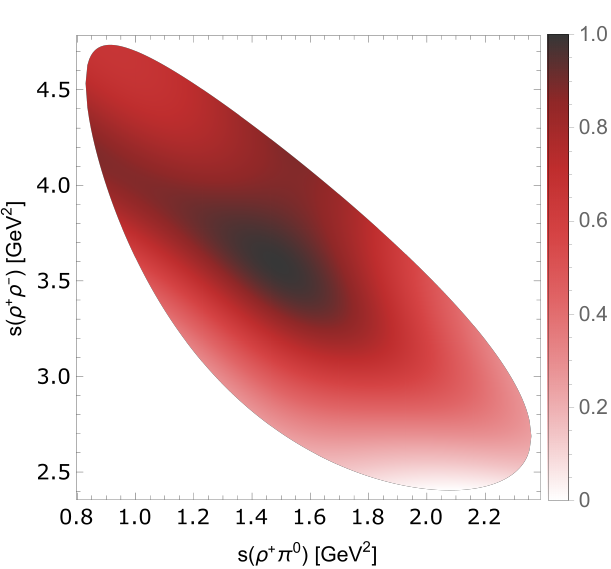}\qquad\includegraphics[width=.42\textwidth]{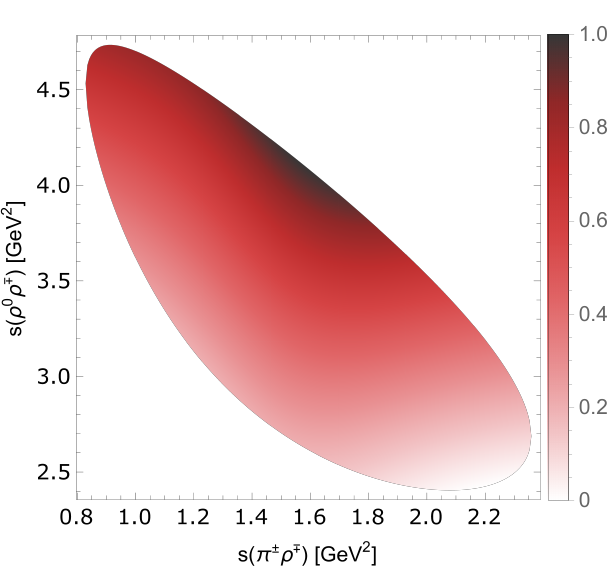}}
        \caption{Dalitz plots for the three body decay $G_{PV}\to\rho\rho\pi$}
        \label{fig:PseudoVectorDalitz}
\end{figure}

\section{Conclusion and Discussion}
\label{sec:concl}

In this paper we have completed our previous study \cite{Hechenberger:2023ljn} of radiative and purely hadronic decays of glueballs in the WSS model by
investigating the decay modes of spin-$1^{\pm-}$ glueballs.
We have found that the latter are dominated
by anomalous vertices involving the Levi-Civita symbol which are uniquely determined by the Chern-Simons action of the flavor branes.

In the case of the vector glueball, such anomalous decays have previously been studied by
Giacosa et al.\ \cite{Giacosa:2016hrm}, however
in the form of just one candidate term among
others which are nonanomalous. While Ref.\ \cite{Giacosa:2016hrm} also obtained $a_1\rho$ decays as dominant anomalous decay, the branching ratio for vector-axial vector decay modes is very much higher in the WSS prediction. For pseudovector glueballs we instead found a dominance of $\rho\pi$.

The WSS model also has direct vertices for the spin-1 glueballs with two vector mesons together with one pseudoscalar. In the case of the $a_1\rho$ channel we found that $\rho\rho\pi$ interferes strongly and negatively with $a_1\rho\to\rho\rho\pi$. Whereas for vector glueballs $a_1\rho$ has a much larger amplitude,
for pseudovector glueballs it is below the direct $\rho\rho\pi$ channel. In Fig.~\ref{fig:VectorDalitz} and \ref{fig:PseudoVectorDalitz} we display the corresponding Dalitz plots for the two spin-1 glueballs with mass given by the WSS model, which shows that in the case of the vector glueball, the resonant decay via $a_1$ should be visible. A clearer signal can however be expected for the decay channels
$G_V\to K_1(1400)K^*$ and $G_V\to K_1(1270)K^*$
which arise in proportion to their $K_{1A}$ content,
since the strange axial vector mesons are more narrow resonances. When the vector glueball mass is extrapolated from the WSS model mass to the prediction of lattice QCD, $K_1(1400)K^*$ becomes
the leading mode.

The decay pattern of the vector glueball 
is thus conspicuously dominated by $a_1\rho$ and $K_1K^*$, which could help in finding its signatures
in reactions such as those studied in \cite{BaBar:2022ahi}
but also implies that a mixing of $J/\psi$ 
with the vector glueball as proposed in \cite{Freund:1975pn,Hou:1982kh,Brodsky:1987bb,Chan:1999px,Hou:1996kw,Hou:1997it} cannot
explain the $\rho\pi$ puzzle in $J/\psi$ and $\psi'$ decays.

We have also revisited the decay pattern of the pseudovector glueball  of Ref.~\cite{Brunner:2018wbv}, 
confirming the conclusion of a very broad
resonance, but correcting the result from
$\Gamma/M\sim 0.92\dots1.37$ to 0.64\dots0.94.
The heavier vector glueball has turned out to be
only slightly less broad, with $\Gamma/M\sim 0.45\dots0.60$. The large widths probably make both spin-1 glueballs
difficult to detect. On the other hand,
their interactions, which are strongly dominated\footnote{Nonanomalous pseudovector glueball interactions from quartic terms in the DBI action have been worked out in Ref.~\cite{Brunner:2018wbv}, where they were found to be negligibly small, suppressed by an extra inverse 't Hooft coupling as well as small coefficients.} 
by anomalous vertices and which are numerically
large, point to an important role
in applications like those studied in
Ref.~\cite{Hechenberger:2024abg}.
% such as threshold
% production of $\eta_{b,c}$ 

\begin{acknowledgments}
We would like to thank Claude Amsler and Francesco Giacosa for useful discussions. F.~H.\ and J.~L.\ have been supported by the Austrian Science Fund FWF, project no. P 33655-N and the FWF doctoral program
Particles \& Interactions, project no. W1252-N27.
\end{acknowledgments}

\bibliographystyle{bib/JHEP}
\bibliography{bib/references}

\end{document}